\documentclass[aps,prx,amsmath,amssymb,floatfix,twocolumn,superscriptaddress,nofootinbib,10pt]{revtex4-2}

\usepackage{graphicx}
\usepackage{dcolumn}
\usepackage{bm}
\usepackage{hyperref}
 \usepackage{amsmath} 
\usepackage{amssymb} 
\usepackage{enumitem} 
\usepackage{hyperref}
\usepackage{color} 
\usepackage{comment}
\usepackage{braket}
\usepackage{soul}
\usepackage{booktabs}
\usepackage{tabularx}
\usepackage{array}
\usepackage{ragged2e}



\begin{document}

\title{Cascade of topological phase transitions and revival of topological zero modes in imperfect double helical liquids 
}

\author{Anna Ohorodnyk}
\affiliation{Institute of Physics, Academia Sinica, Taipei 115201, Taiwan}
\affiliation{Department of Electronics and Electrical Engineering, National Yang Ming Chiao Tung University, Hsinchu 30010, Taiwan}

\author{Chen-Hsuan Hsu}
\affiliation{Institute of Physics, Academia Sinica, Taipei 115201, Taiwan}
\affiliation{Physics Division, National Center for Theoretical Sciences, Taipei 106319, Taiwan}

\date{\today}

\begin{abstract}

Two parallel helical edge channels hosting interacting electrons, when proximitized by local and nonlocal pairings, can host time-reversal-invariant pairs of topological zero modes at the system corners. Here we show that realistic imperfections substantially enrich the physics of such proximitized double helical liquids. Specifically, we analyze this platform and its fractional counterparts in the presence of pairing and interaction asymmetries between the two channels, as well as random spin-flip terms arising from either magnetic disorder or coexisting charge disorder and external magnetic fields. Using renormalization-group analysis, we determine how Coulomb interactions, pairings, and magnetic disorder collectively influence the transport behavior and topological properties of the double helical liquid. As the system now belongs to class BDI, an additional topological phase supporting a single Majorana zero mode per corner emerges. We further show how additional pairing or Coulomb asymmetry influences the stability of various topological phases and uncovers a revival of Majorana zero modes and cascades of successive topological transitions through topological phases characterized by a $\mathbb {Z}$ invariant, which are accessible through controlling the electrical screening effect. We also analyze the spatial structure of the zero modes and the bulk gap closing through channel-resolved density profiles, revealing observable features through scanning tunneling microscope. In contrast to conventional understanding, disorder is not merely detrimental, as it in general allows for a tuning knob that qualitatively reshapes the topological superconductivity in imperfect helical liquids.
  
\end{abstract}

\maketitle

\section{\label{sec:level1} Introduction }
 
Topological superconductivity represents a quantum phase in which superconductivity coexists with topologically protected excitations. In one-dimensional systems, such phases can give rise to topological zero modes with non-Abelian statistics~\cite{Nayak:2008,Alicea:2012,Beenakker:2013,DasSarma:2015,Sato:2017,Beenakker:2020}. Majorana zero modes serve as the canonical example of such excitations and are central to proposals for topological quantum computation~\cite{Ivanov:2001,Nayak:2008,Sau:2010,Beenakker:2013,DasSarma:2015,Beenakker:2020}.
A widely studied class of realizations in proximitized Rashba nanowires~\cite{Sato:2009,Sato:2009b,Oreg:2010,Lutchyn:2010,Klinovaja:2012a,Alicea:2012,Beenakker:2013,DasSarma:2015,Frolov:2020,Sato:2017,Das:2012,Deng:2012} 
relies on external magnetic fields to break time-reversal symmetry, but this requirement unavoidably introduces experimental difficulties such as orbital depairing and the suppression of superconductivity.

Among various alternatives, such as atomic chains and other low-dimensional platforms~\cite{Kitaev:2001,Kitaev:2003,Wong:2012,Leijnse:2012,Keselman:2013,Klinovaja:2013a,Nakosai:2013,Zhang:2013,Haim:2014,Klinovaja:2014b,Hsu:2015,Hoffman:2016b,Schrade:2017,Thakurathi:2018,Dvir:2023},  
a promising route for avoiding external magnetic fields is to engineer superconductivity in helical liquids~\cite{Klinovaja:2014helical,Hsu:2018,Schulz:2019,Hsu:2021,Hung:2025}. Helical liquids arise at the edges or hinges of time-reversal-invariant topological insulators~\cite{Kane:2005a,Kane:2005b,Bernevig:2006,Bernevig:2006b,Wu:2006,Xu:2006,Hasan:2010,Qi:2011,Hsu:2021,Noguchi:2021,Jia:2022,Wang:2023,Weber:2024,Yu:2024,Hsu:2025}, 
providing a  platform with time-reversal-invariant one-dimensional modes that are well suited for realizing topological and correlated quantum states.
When brought into proximity with superconductors, such helical channels can host Majorana and parafermion zero modes~\cite{Fu:2008,Fu:2009,Tanaka:2009,Crepin:2014a,Crepin:2014b,Klinovaja:2014helical,Crepin:2015,Haim:2016,Hsu:2018,Yan:2018,Fleckenstein:2019,Haim:2019,Hsu:2021,Fleckenstein:2021,Keidel:2018,Novik:2020,Zhang:2020}, 
providing time-reversal-invariant settings for stabilizing and detecting zero modes, as well as forming topological quantum bits~\cite{Liu:2014,Schrade:2018,Schrade:2022}.
Recent developments have further extended the platform of helical liquids to moiré and twisted bilayer systems~\cite{Kang:2024a,Kang:2024b}, where fractional analogues of quantum Hall edges can emerge~\cite{Kang:2024a}. These systems open new opportunities for realizing correlated and topological phases.
 
Despite these advances, realistic devices rarely exhibit ideal helical transport. Experiments consistently report deviations from perfectly quantized conductance~\cite{Konig:2007,Konig:2008,Roth:2009,Knez:2011,Gusev:2011,Grabecki:2013,Suzuki:2013,Gusev:2014,Knez:2014,Olshanetsky:2015,Suzuki:2015,Fei:2017,Bendias:2018,Wu:2018,Gusev:2019,Culcer:2020},
indicating the presence of backscatterings in practical platforms~\cite{Gusev:2019,Culcer:2020,Hsu:2021} and triggering numerous studies on possible mechanisms~\cite{Maciejko:2009,Jiang:2009,Strom:2010,Hattori:2011,Lunde:2012,Budich:2012,Lezmy:2012,Schmidt:2012,Delplace:2012,Crepin:2012,DelMaestro:2013,Altshuler:2013,Vayrynen:2013,Geissler:2014,Kainaris:2014,Vayrynen:2014,Baum:2015,Chou:2015,Yevtushenko:2015,Fleckenstein:2016,Vayrynen:2016,Xie:2016,HsuLocalization:2017,Kharitonov:2017,Wang:2017,Groenendijk:2018,Muller:2017,HsuTransport:2018,Vayrynen:2018,Dietl:2023a,Dietl:2023b}. 
Such imperfections can originate from disorder or impurities that are inherently present in nanostructures and make it essential to understand how they influence the robustness of topological zero modes under superconducting proximity. In fact, this issue is subtle: 
in the absence of superconductivity, breaking the time-reversal symmetry is insufficient to affect helical-edge transport, and spin-nonconserving scattering processes are required to produce elastic backscatterings~\cite{Tanaka:2011,Eriksson:2012,Eriksson:2013}. 
Furthermore, strong disorder and interactions can qualitatively alter the boundary phases of two-dimensional time-reversal-invariant topological superconductors~\cite{Son:2024}.
Concerning single helical edges, the interplay among electron-electron interactions, Zeeman fields and disorder has been analyzed, revealing rich phase diagrams where disorder competes with superconductivity~\cite{Bakhshipour:2025a,Bakhshipour:2025b}. However, a system consisting of coupled helical liquids introduces new degrees of freedom that fundamentally alter this landscape. 

When the two helical channels are integrated with superconductors, additional sources of imperfections arise, including asymmetric pairing induced by nonuniform proximitization between the channels and differences in the Coulomb interaction strength of the two channels caused by the local screening environment. These effects are conventionally viewed as detrimental, since they modify the low-energy behavior of helical channels and might reduce the stability of Majorana zero modes. Indeed, theoretical studies have shown that electron-electron interactions and electron-phonon coupling~\cite{HsuPhonons:2024} can weaken and even destabilize the zero modes in double helical liquids, in parallel to the studies on electron-electron interactions~\cite{Loss:2011}, electron-phonon coupling~\cite{Aseev:2019}, and disorder~\cite{Thakurathi:2018} in nonhelical platforms.

The central message of this work is that these imperfections are, in fact, not merely obstacles. When treated on equal footing with interactions and superconducting pairing, they introduce new mechanisms that reorganize the topological landscape of proximitized double helical liquids, enabling phases and transitions that do not appear in the idealized, perfectly symmetric limit.

In this work, we investigate a system of two interacting helical channels with both intrachannel and interchannel pairings, and introduce imperfections commonly present in realistic devices, including  random spin-flip backscattering as well as pairing and Coulomb asymmetries between the two channels.
These ingredients arise naturally from magnetic impurities, in-plane fields combined with charge disorder, and inhomogeneous local environment of the two channels. We derive an analytical expression for the number of topological zero modes as a function of system parameters, and combine it with the renormalization-group (RG) framework. We show that spin-flip backscattering generically detunes the two zero-mode conditions that are degenerate in the clean limit, opening a new phase where the system supports a single Majorana zero mode. 
We also demonstrate that interactions and disorder can drive realistic systems into this regime even when the clean limit remains topologically trivial. We further find that backscattering enhances the effects of weak pairing asymmetry, leading to a controllable revival of topological zero modes and a cascade of transitions between phases labeled by a $\mathbb Z$ invariant. 
Finally, we analyze the channel-resolved density profiles, which reveal observable features in scanning probes upon varying the system parameters. 
Taken together, these results show how realistic imperfections reorganize both the topological and transport properties of (fractional) helical liquids, with electrically tunable features in the platform.

The rest of this article is organized as follows.
In Sec.~\ref{sec:Hamiltonian}, we introduce our setup and the corresponding bosonized Hamiltonian, consisting of interacting electrons in two helical channels with proximity-induced local and nonlocal pairings and random spin-flip backscatterings. 
In Sec.~\ref{sec:rg_flow}, we discuss the RG flow equations and representative RG flow diagrams. 
In Sec.~\ref{sec:transport}, we explore the transport properties of the system. 
We use the renormalized couplings to analyze how various phases evolve in Sec.~\ref{subsec:transport_PD}, and compute the localization length and temperature in the insulating phase in Sec.~\ref{subsec:localization}.  
In Sec.~\ref{sec:topological_properties}, we characterize the system according to its topological properties and the associate zero modes.
We discuss the formula of the number of the zero modes and examine how various ingredients existing in imperfect helical liquids influence the topology in Sec.~\ref{subsec:Majorana_criterion}.
Combining with RG analysis, we obtain topological phase diagrams in Sec.~\ref{subsec:topo_PD}. 
In Sec.~\ref{sec:3dphasediagrams}, we construct both the transport- and topology-based phase diagrams in three-dimensional parameter spaces. In Sec.~\ref{sec:microscopic_features}, we visualize the spatial evolution of the Majorana zero modes and the bulk gap closing using channel-resolved density profiles.
Finally, in Sec.~\ref{sec:discussion}, we summarize our main findings and discuss possible extensions and experimental realizations. 
Technical details are given in the appendices. 
The experimentally relevant platforms and their material parameters are summarized in Appendix~\ref{Appendix:platforms}.
The derivation of random spin-flip backscattering terms are given in Appendix~\ref{Appendix:rs_terms}.
The derivation of the RG flow equations, including the generalization to fractional helical liquids, is in Appendix~\ref{Appendix:RG}. 
The symmetry analysis of the effective model, solutions of the corresponding Bogoliubov-de Gennes equation and criterion of the zero modes are collected in Appendix~\ref{Appendix:single-particle-analysis}. 
Additional details of the numerical analysis are provided in Appendix~\ref{Appendix:more}.

\begin{figure}[t]
\centering
    \includegraphics[width=1.0\linewidth]{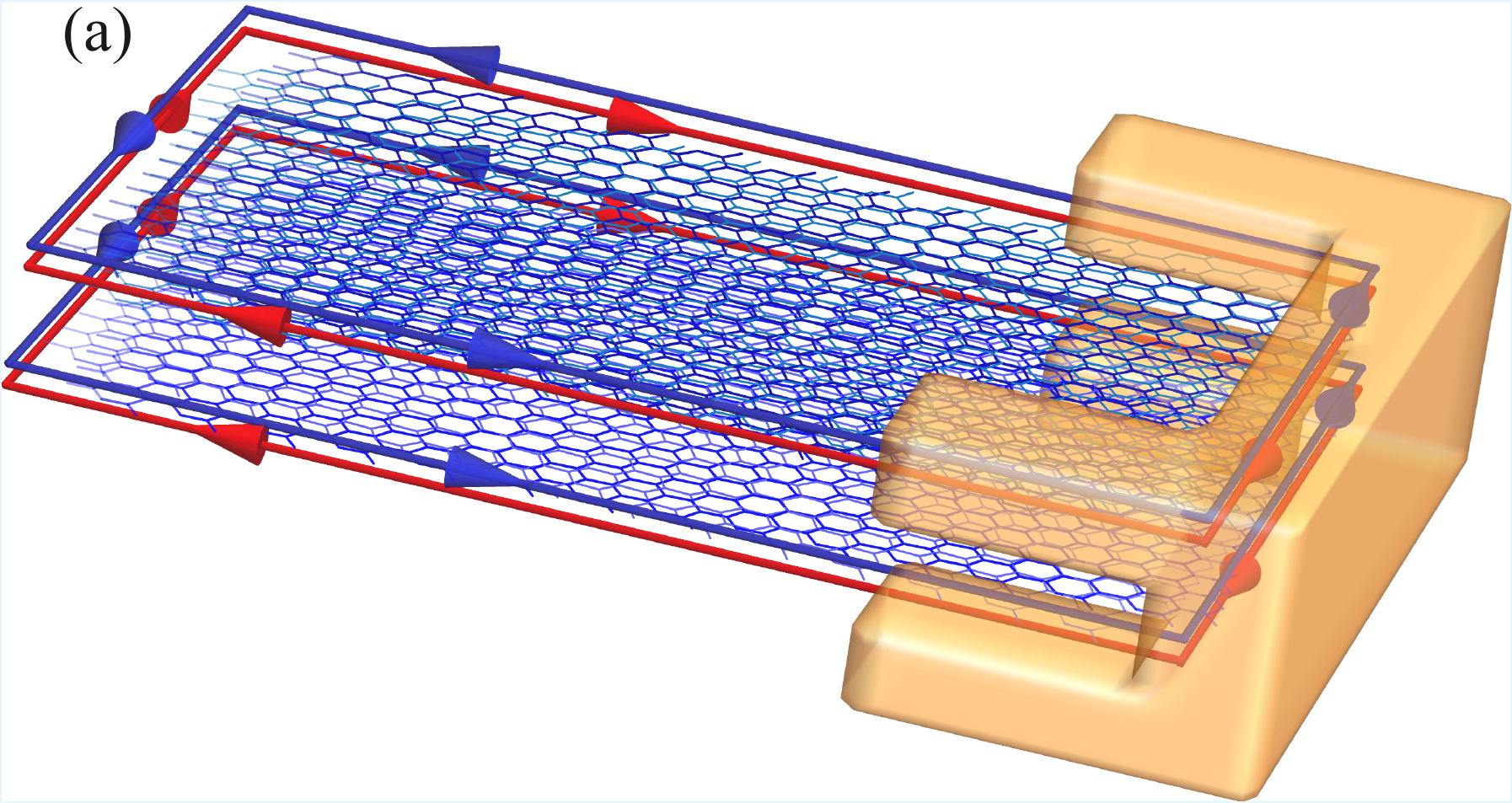}\\[0.1em]
    \includegraphics[width=1.0\linewidth]{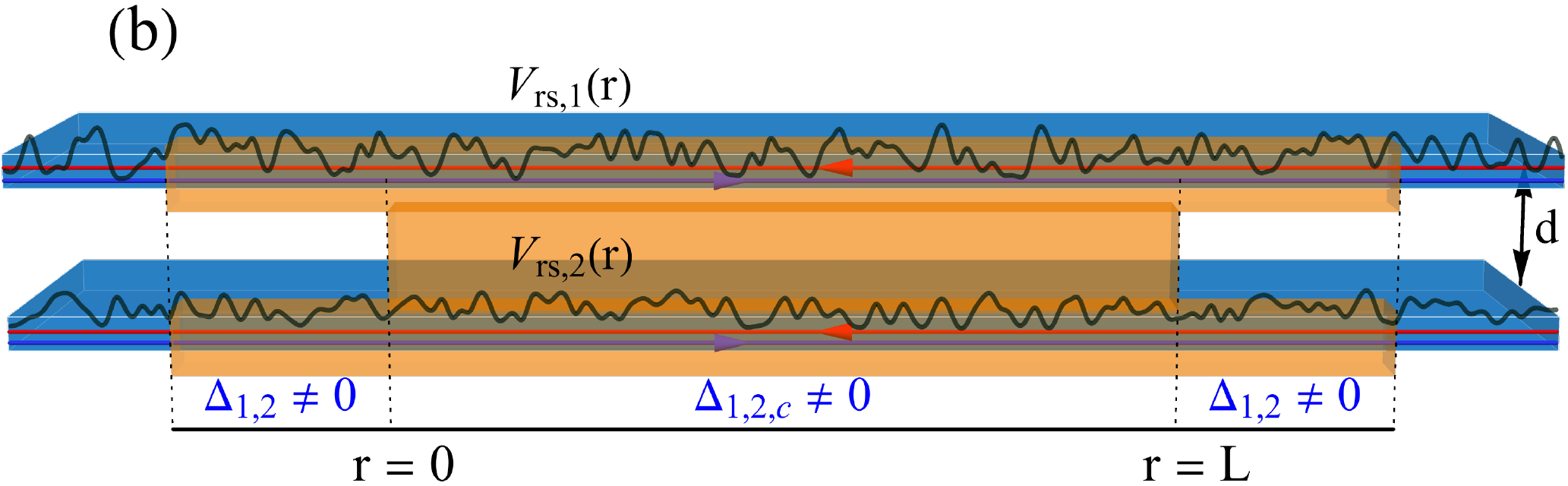}
    \caption{Schematic of a proximitized double helical liquid consisting of two parallel edge channels of time-reversal-invariant topological insulators (blue), separated by a distance $d$. The channels are in contact with an $s$-wave superconductor (orange). 
    (a) Possible realization based on twisted bilayer structures. 
    (b) Side view of the setup along the channel coordinate $r$. Both local and nonlocal pairings are induced in the edge region $r\in[0,L]$, and only local pairing occurs for $r<0$ and $r>L$.  
    The edges are subject to spin-flip backscatterings with strength $V_{{\rm rs}, n}$. 
    }
    \label{fig:setup}
\end{figure}

\section{Setup and model } 
\label{sec:Hamiltonian}

We consider the setup illustrated in Fig.~\ref{fig:setup}, which includes two parallel topological edge channels hosting right-moving down-spin and left-moving up-spin modes. 
For experimentally relevant parametes, such as edge-state velocities, their wave function decay lengths, and bulk gaps of candidate quantum spin Hall materials, we refer to  Appendix~\ref{Appendix:platforms}.
To describe the interacting electrons in the (fractional) helical channels, we express the electron fields, $\psi_{n} \approx e^{i k_F r} R_{\downarrow,n} (r) + e^{-i k_F r} L_{\uparrow,n}  (r)$, as 
\begin{subequations}
\label{fermionic_fields_in_bosonic}
\begin{align}
R_{\downarrow,n}(r) =& \frac{U_{R,n}}{\sqrt{2 \pi a}}e^{im[-\phi_n(r) + \theta_n(r)]}, \\[2ex]
L_{\uparrow,n} (r) =& \frac{U_{L,n}}{\sqrt{2 \pi a}}e^{im[\phi_n(r) + \theta_n(r)]}, 
\end{align}
\end{subequations}
with the coordinate $r$ along the channels, Klein factors $U_{R,n}$, $U_{L,n}$, Fermi wave vector $k_{F}$ and short-distance cutoff $a$, taken to be the transverse decay length of the edge states.
The bosonic fields,  $\phi_n$ and $\theta_n$, satisfy 
\begin{equation}
\label{fields_commutation_relation}
[\phi_n(r),\theta_{n'}(r')]=\frac{i\pi}{2m} \delta_{nn'}\text{sign}(r'-r).
\end{equation}
With an odd $m$, the system corresponds to a time-reversal-invariant generalization of Laughlin states at  $\nu = 1 / m$ fillings~\cite{Klinovaja:2014helical,Levin:2009,Neupert:2011,Santos:2011}.  

With the introduced bosonic fields, we construct the Hamiltonian as
\begin{equation}
H = H_{\text{el}} + H_{\rm s} + H_{ \times} + H_{\rm rs } .
\end{equation}
The first term describes double helical (Tomonaga-Luttinger) liquids formed in two parallel helical channels labeled by the index $n \in \{ 1, 2 \} $,
\begin{equation}
H_{\text{el}} = \sum_{n \in \{1,2\} } \int \frac{\hbar  dr}{2\pi}
\left[ u_n K_n \big(\partial_r \theta_n\big)^2 + \frac{u_n}{K_n} \big(\partial_r \phi_n\big)^2 \right],
\end{equation}
where $\phi_n$ and $\theta_n$ denote the bosonic dual fields with the velocities 
$u_n \equiv v_{ {\rm F} }/(K_{n} m)$  
and interaction parameter $K_n$. 
In our convention, the value $K_n = 1$ corresponds to the noninteracting limit, whereas $K_n < 1$ ($K_n > 1$) indicates repulsive (attractive) electron-electron interactions. 
It has been shown that the interaction strength in the edge channels can be electrically tuned by  gates~\cite{Wang:2023}.  

The intrachannel, local pairing term,
\begin{equation}
H_{\rm s} = \sum_{n \in \{1,2\} } \frac{\Delta_n}{\pi a} \int dr \, \cos \left[2 m \theta_n(r)\right],
\end{equation}
represents conventional $s$-wave pairing induced independently in each channel via the proximity effect, with the corresponding pairing strength $\Delta_{n}$ extending also into the region of $r<0$ and $r>L$; see Fig.~\ref{fig:setup}.
On the other hand, the interchannel, nonlocal pairing contribution,
\begin{align}
H_{\times} = & \frac{2\Delta_c}{\pi a} \int dr \,
\cos \left[m(\theta_1(r) + \theta_2(r))\right] \nonumber  \\
& \times
\cos \left[m(\phi_1(r) - \phi_2(r))\right],
\end{align}
corresponds to crossed Andreev pairing, in which a Cooper pair from the parent superconductor splits and its constituent electrons tunnel into different helical channels with the pairing strength $\Delta_{c}$ within the range $r \in [0,L]$, as illustrated in Fig.~\ref{fig:setup}. Microscopically, this term arises from the same superconducting proximity effect that generates the local intrachannel pairing and can be obtained by integrating out the superconducting degrees of freedom to second order in the interface tunneling~\cite{Hsu:2018}. Its bare amplitude depends on the tunnel couplings between the superconductor and the two channels and on their spatial separation. It is therefore appreciable when the interchannel separation is comparable to or smaller than the coherence length of the parent superconductor.

Finally, random spin-flip backscattering can be included as 
\begin{equation}
\label{rs_hamiltonian}
H_{\rm rs} =
\sum_{n} \int \frac{dr}{2\pi a} \,
\left[
V_{{\rm rs}, n}(r) e^{ 2m i \phi_n (r)} + \text{H.c.}
\right] ,
\end{equation}
where $V_{{\rm rs}, n}(r)$  represents random spin-flip backscattering potential in channel $n$. The above term breaks spin-momentum locking and introduces backscattering characterized by~\cite{HsuLocalization:2017, HsuTransport:2018} 
\begin{equation}
   \overline{\langle V_{{\rm rs}, n}^\dagger(r) V_{{\rm rs}, n'}(r') \rangle} = D_{n}  \delta_{nn'} \delta (r-r'), 
\end{equation}
where the overbar denotes averaging over disorder realizations and $\braket\dots$ denotes the quantum expectation value for a fixed configuration. 
The  effective backscattering strength is related to $V_n$ as $D_n = a V_{ n}^2$,   
with the root-mean-square amplitude  $V_{n} = \big[ \overline{\langle |V_{{\rm rs}, n}(r)|^2 \rangle} \big] ^{1/2}$ of the random potential; further details are provided in Appendix~\ref{Appendix:rs_terms}.
As a remark, the physical mechanisms generating spin-flip backscattering can also suppress the proximity-induced pairing. 
We therefore restrict our analysis to the regime in which these perturbations remain below the pair-breaking threshold of the parent superconductor and the induced pairing remains finite. The pairing strengths $\Delta_n$ introduced above should be understood as effective values that already incorporate such microscopic suppression. 

To proceed, we perform the replica method to average over the random potential~\cite{Giamarchi:2004}, which allows us to derive the corresponding contribution to the effective action. This procedure generates an additional term in the effective imaginary-time action, 
\begin{equation}
\label{eq:rs_action}
\begin{aligned}
 &   \frac{\delta S_{\rm rs}}{\hbar}=-\sum _n \frac{ D_n}{(2\pi \hbar a)^2}
    \\
&    \times 
    \int_{ u |\tau-\tau'|>a} dr d\tau d\tau' 
    \cos[2m\phi_n(r,\tau) -2m\phi_n(r,\tau')].
\end{aligned}
\end{equation}

Before proceeding, we briefly remark on the topological properties of the system in the clean limit. When random spin-flip backscattering is absent, the system enters a topologically nontrivial phase and develops a twofold ground-state degeneracy when the nonlocal pairing dominates over the local one~\cite{HsuPhonons:2024}. With the bosonization, it can be shown that this degeneracy is protected by the conservation of fermion parity associated with the spin difference between the two helical channels~\cite{HsuPhonons:2024}, extending the earlier analysis in the single-particle regime~\cite{Klinovaja:2014helical}.

Below, we investigate the system in the presence of the spin-flip backscattering term $H_{\mathrm{rs}}$, which modifies effective pairing strengths and consequently influences both transport properties and topological stability of the Majorana zero modes.

\section{RG flow equations and flow diagrams } 
\label{sec:rg_flow}

To investigate the stability of various electronic phases in the  system, in this section we derive the RG flow equations to examine the relevance of each of the non-quadratic terms in the Hamiltonian. 
This allows us to extract the coupling constants renormalized under the RG flow. 

To proceed, we introduce the following dimensionless couplings, 
\begin{equation}
\begin{aligned}
\tilde \Delta_n \equiv \frac{\Delta_n a}{\hbar u_n}, \;
\tilde \Delta_c \equiv \frac{\Delta_c a}{\hbar \sqrt{u_1 u_2}}, \; 
\tilde D_n\equiv \frac{2 a^2 V_{ n}^2}{\pi \hbar^2 u_n^2}, 
\end{aligned}
\end{equation}
corresponding to local pairing ($ \tilde\Delta_n $), nonlocal pairing ($ \tilde \Delta_c $), and backscattering  ($\tilde{D}_n$) strengths. 
For convenience, we also introduce $\tilde \Delta_\pm \equiv (\tilde \Delta_1 \pm \tilde \Delta_2) / 2$ and 
 $\tilde D_\pm \equiv (\tilde D_1 \pm \tilde D_2) / 2$.

\begin{figure}[t]
\centering
    \includegraphics[width=1.0\linewidth]{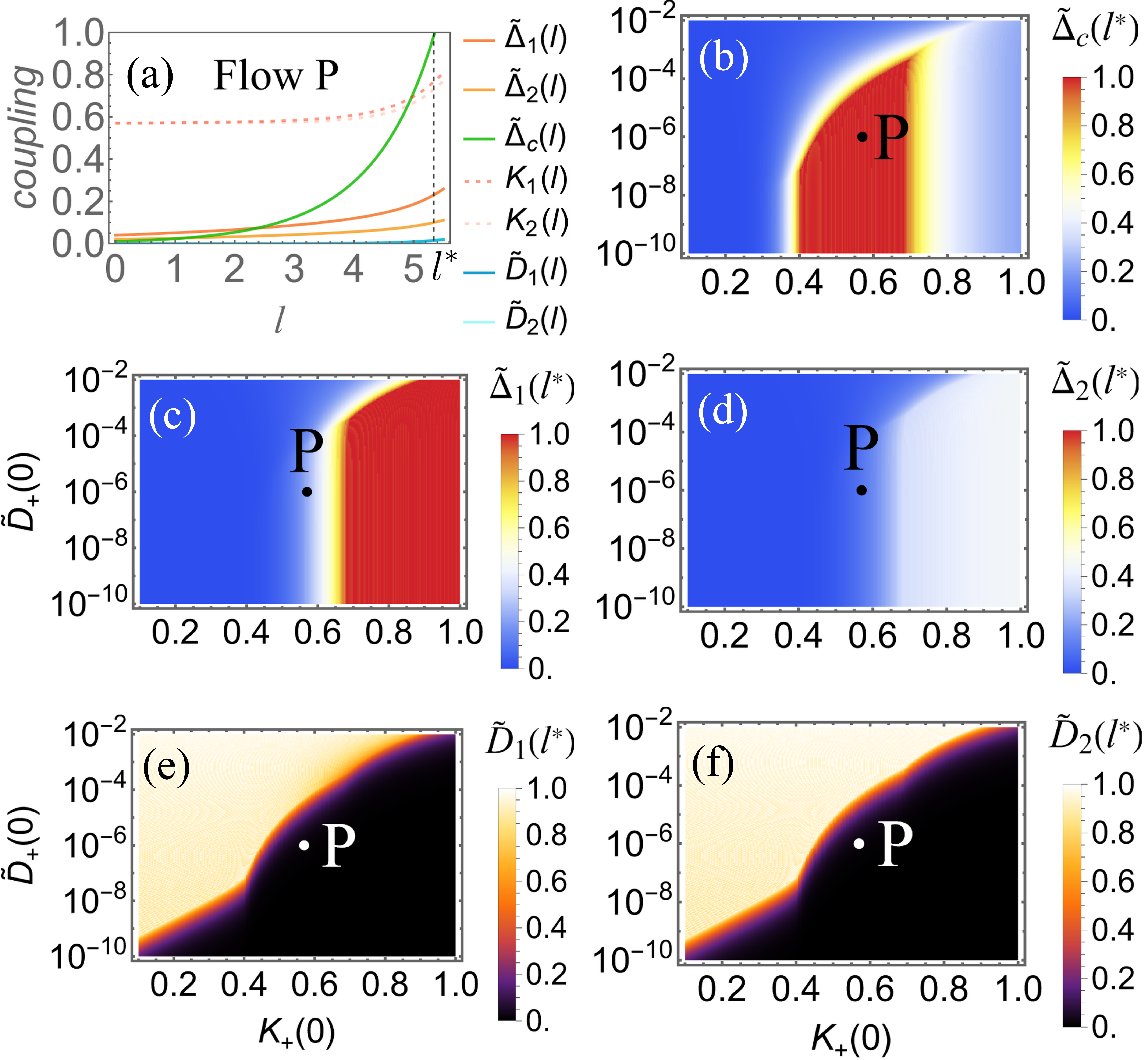}
    \caption{RG flow and renormalized coupling strengths with $\tilde \Delta_+(0) =0.03$, $\tilde \Delta_-(0) = 0.01$, $\tilde\Delta_c(0)=0.01$ and $\tilde D_-(0) =K_-(0) =0$. (a) RG Flow for the initial parameter set P with $\tilde D_+(0) = 10^{-6}$ and $ K_{+} (0) = 0.57 $.
    The label $l^*$ marks the cutoff scale where the flow stops. 
    (b--f) Color maps of the renormalized couplings in the $[K_+(0),\tilde D_+(0)]$ plane for (b--d) the pairing strengths $\tilde \Delta_{n}(l^*)$ and $\tilde \Delta_{c}(l^*)$ and  (e--f) the backscattering strengths $D_{n}(l^*)$. The marked dot P corresponds to the parameter set used in Panel~(a). See Table~\ref{Table:Parameters} for the complete set of the adopted parameter values. 
   }
    \label{fig:pointPflow}
\end{figure}

Upon changing the cutoff, the evolution of the dimensionless coupling is governed by a set of coupled differential equations. 
Following the algebra presented in Appendix~\ref{Appendix:RG}, we derive the RG flow equations with the dimensionless length scale  $l$,  
\begin{subequations}
\begin{align}
&\frac{d\tilde{\Delta}_n}{dl} = \left(2 - \frac{m}{K_n}\right) \tilde{\Delta}_n, \\
&    \frac{d\tilde{\Delta}_c}{dl} = \left[2 - \frac{m}{4}\left(K_1 + K_2 + \frac{1}{K_1} + \frac{1}{K_2}\right)\right] \tilde{\Delta}_c, \\
&    \frac{d\tilde{D}_n}{dl} = (3 - 2mK_n) \tilde{D}_n, \\
\label{eqn:rg_flow_K}
&    \frac{dK_n}{dl} = m\left[\tilde{\Delta}_n^2 + \frac{1}{2}(1 - K_n^2)\tilde{\Delta}_c^2 - \frac{K_n^2}{2}\tilde{D}_n\right], \\
&    \frac{du_n}{dl} = -m\frac{u_n K_n}{2} \tilde D_n ,
    \end{align}
\end{subequations}  
where we include the leading-order contributions.

\begin{figure*}[ht]
\centering
\includegraphics[width=1.0\linewidth]{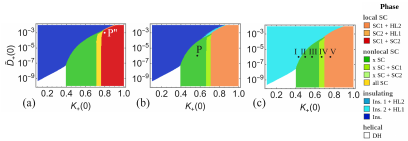}
\caption{Phase diagrams based on the  transport properties in the $[K_+(0),\,\tilde D_+(0)]$ plane. 
We consider (a) the symmetric-channel case with $\tilde \Delta_-(0) = K_-(0) =0$, (b) the pairing asymmetry case with $\tilde \Delta_-(0) = 0.01$, and (c) the Coulomb asymmetry case with $K_-(0)=0.05$.
We identify the dominant (and subdominant) phases among local superconductivity in channel $n \in \{1,2\}$ (labeled as ``SC $n$''), nonlocal superconductivity ($\times$ SC), insulating channel $n$ (Ins. $n$), helical liquid in channel $n$ (HL $n$), double helical liquid (DH), fully insulating channels (Ins.), and superconductivity with comparable local and nonlocal pairings (all SC). 
See Table~\ref{Table:Parameters} for the complete set of the adopted parameter values.  
}
\label{fig:PD_transport1}
\end{figure*}

The parameter values adopted in the numerical analysis are chosen to span representative ranges for semiconductor-based quantum spin Hall materials (Table~\ref{Tab:semiconductors}), monolayer materials (Table~\ref{Tab:monolayer}) and twisted nanostructures (Table~\ref{Tab:moire}). 
Specifically, we fix the Fermi velocity $v_F=10^5~{\rm m/s}$ and the short-distance cutoff $a(0)=5~{\rm nm}$, where the latter effectively represents the transverse decay length of the edge states. This choice yields a characteristic energy cutoff of $O(10$~meV). Scaled by this bandwidth, the dimensionless coupling $\tilde \Delta_+(0)$ or $\tilde \Delta_c (0)=0.01$ corresponds to a physical proximity-induced gap of O($ 0.1$~meV).

Given a set of initial parameters defined at a microscopic length scale (see Table~\ref{Table:Parameters} in Appendix~\ref{Appendix:more}), we numerically solve the RG flow equations.
The integration proceeds until one of the dimensionless couplings flows to unity, or until a maximum physical length scale  $ \ln(L_{\rm ch}/a)$ set by the system size $L_{\rm ch}$ or alternatively $ \ln(L_{\rm th}/a)$ based on the thermal length $L_{\rm th} = \hbar v_F / (k_B T)$.
In what follows, we assume identical backscattering strength in the channels, $\tilde D_{-}(0)=0 $; see Appendix~\ref{Appendix:RG_disorder_asymmetry} for results beyond this assumption. 
Additionally, we focus on the numerical analysis of the quantum spin Hall edges ($m=1$), as the  fractional edges ($m>1$) tend to suppress the pairing and lead to more straightforward consequences; the corresponding results for $m>1$ are summarized in Appendix~\ref{Appendix:m3_fractional}.

To illustrate how RG flow determines which coupling ultimately dominates, in Fig.~\ref{fig:pointPflow} we consider a representative set (labeled as P) in the parameter space. As shown in Fig.~\ref{fig:pointPflow}(a), the nonlocal pairing $\tilde \Delta_c(l)$ grows most rapidly, owing to the corresponding scaling dimension,  and decides the fate of the flow by reaching strong coupling at the cutoff scale, while the local pairings $\tilde \Delta_{n}(l)$ increase only moderately. This behavior is mirrored in the color maps of the renormalized couplings: Fig.~\ref{fig:pointPflow}(b) shows a fully saturated $\tilde \Delta_c(l^*)$ at point P, whereas Figs.~\ref{fig:pointPflow}(c--d) display smaller but noticeable renormalized values of the local pairings, with $\tilde \Delta_1(l^*)>\tilde \Delta_2(l^*)$, consistent with the small initial asymmetry. At this particular point P, the backscattering couplings in Figs.~\ref{fig:pointPflow}(e--f) are not  relevant for the flow, even though they become dominant in  regions where the initial interactions and backscattering are stronger. Thus, point P lies in a regime where nonlocal pairing takes over the RG trajectory, with the local pairings contributing only as weak corrections, and backscattering contributions remaining negligible. 
For completeness, we present additional renormalized coupling diagrams scanning pairing asymmetry at a fixed interaction strength in Fig.~\ref{fig:pointPprimeflow} in Appendix~\ref{Appendix:more}.

As shown above, the RG analysis allows us to deduce the pairing and backscattering strengths for interacting electrons in the double helical channels. Below we examine this in a broader region of the parameter space. 

\section{Transport properties}
\label{sec:transport}

In this section, we explore transport properties based on the RG flow equations derived in the preceding section, which allows us to determine the ultimate fate of the system at low energies. 
In the fully insulating phase where both of the backscattering terms $\tilde D_n$ become the most relevant couplings, we expect exponential localization of electronic states along the channel, in analogous to the Anderson localization in nonhelical systems~\cite{Abrahams:1979, Abrahams:2010}, which will be analyzed below. 

\subsection{Phase diagrams based on transport properties}
\label{subsec:transport_PD}

As demonstrated above, the dominant coupling at the end of the RG flow determines the low-temperature transport characteristics of the double helical channels. Here we carry out  the RG procedure for different choices of the initial couplings and determine which physical process ultimately controls the low-energy behavior. 

The results are summarized in the phase diagrams in Fig.~\ref{fig:PD_transport1}, which illustrate how the interaction strength, pairing asymmetry, and backscattering jointly give rise to various phases including  local superconducting, nonlocal superconducting, insulating, metallic regimes, and their mixtures in the two channels.
For a systematic discussion, we group the phase diagrams into two sets, with Fig.~\ref{fig:PD_transport1} illustrating  how varying interaction and backscattering strengths influence the electronic phases. In Appendix~\ref{Appendix:more}, we additionally present Fig.~\ref{fig:PD_transport2}, which examines the  situation where pairing asymmetry is varied at a fixed interaction strength.

\label{subsubsec:PD_Kplus_Dplus}

We start with a representative phase diagram in Fig.~\ref{fig:PD_transport1}(a)  for a system with two identical channels. In the strongly interacting regime, the system remains as helical liquids at weak backscattering strengths and flows to an insulating phase at strong backscattering strengths. At intermediate interactions, a robust  superconducting phase emerges for sufficiently weak backscattering, where nonlocal pairing $ \Delta_c$ dominates. In the weakly interacting regime, local pairings ($\Delta_{n}$) prevail, leading to a local superconducting phase. An intermediate phase, in which nonlocal and local superconducting orders coexist, emerges near the boundary separating the nonlocal and local superconducting regimes.

Next, we move on to Fig.~\ref{fig:PD_transport1}(b), where a small pairing asymmetry is introduced through a finite $ \Delta_-(0)$, and all the other parameters remain identical to Fig.~\ref{fig:PD_transport1}(a). 
While the overall structure is similar to the symmetric case, the nonlocal superconducting region shrinks because the larger local pairing $\Delta_1$  eventually dominates over a wider range of interaction strengths.
Likewise, at weak interactions, the system enters an asymmetric phase in which one channel becomes superconducting while the other remains metallic. An additional intermediate regime emerges at weak interactions and strong backscattering, where one channel turns insulating while the other stays metallic. 
These changes illustrate how pairing imbalance tips the RG flow toward a single favored channel.

Finally, we examine the case with distinct interaction strengths in the channels, with $K_-(0)=0.05$ and present Fig.~\ref{fig:PD_transport1}(c). The nonlocal superconducting region is again reduced as compared to Fig.~\ref{fig:PD_transport1}(a), since the channel with the weaker interactions now gains a slight advantage and its local pairing stays dominant over a wider range of $K_+(0)$. At weak interactions, this produces an asymmetric local superconducting phase, analogous to  Fig.~\ref{fig:PD_transport1}(b). At strong backscattering strengths, the fully insulating state is replaced by an asymmetric one in which one channel becomes an insulator while the other remains gapless. At even stronger interactions, the mixed phase with coexisting insulating and metallic channels broadens.  
Overall,  Fig.~\ref{fig:PD_transport1} shows that asymmetries in the pairing or interaction strengths can lead to phase transitions and additional phases;
see Appendix~\ref{Appendix:more}  for additional phase diagrams varying different parameters. 
Crucially, this will result in topological phase transitions, as will be discussed in Sec.~\ref{sec:topological_properties}.

\begin{figure}[t]
\centering
    \includegraphics[width=1.0\linewidth]{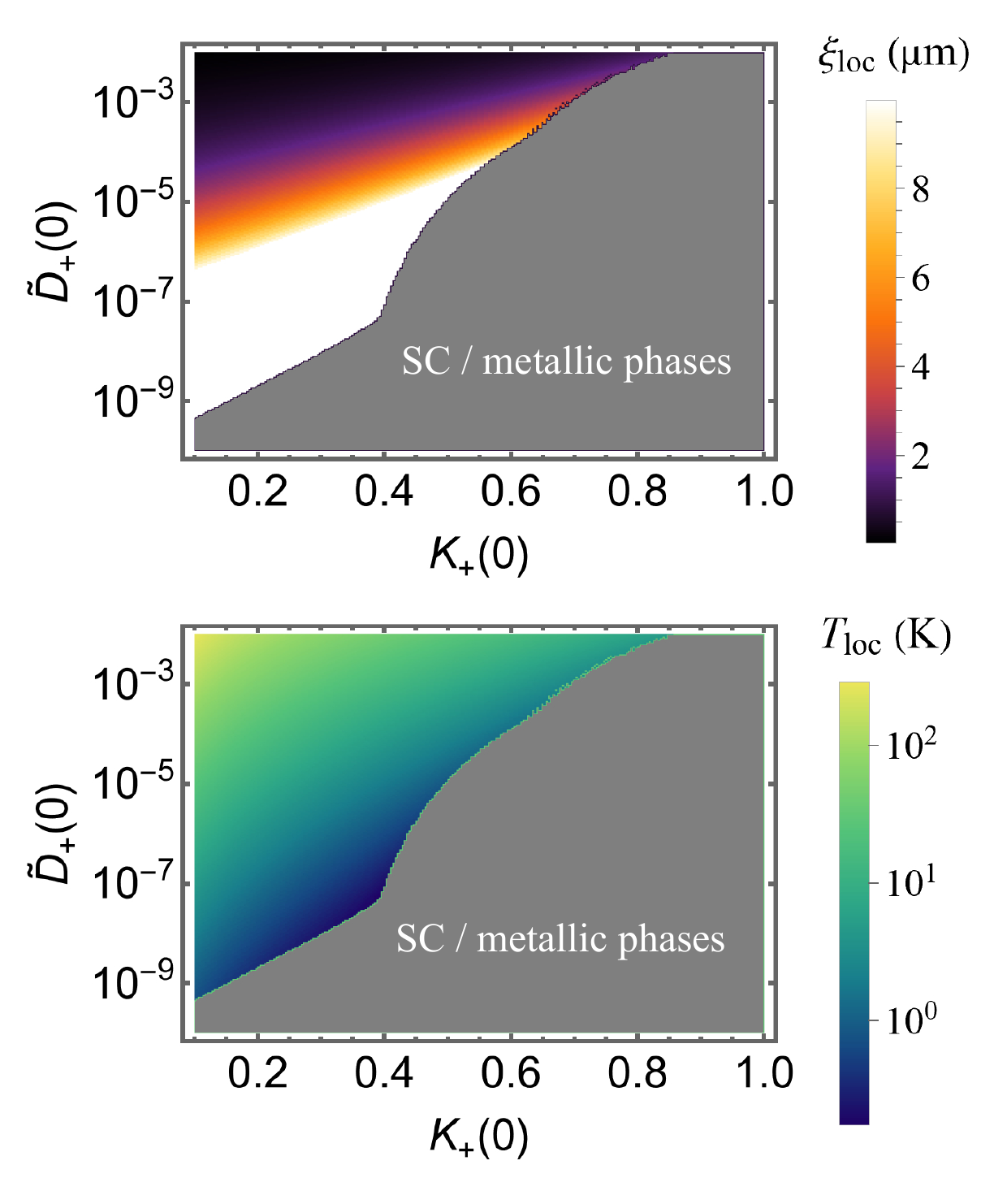}
    \caption{Localization length ($\xi_{\rm loc}$) and temperature ($T_{\rm loc}$, on logarithmic scale) within the fully insulating phase in the parameter space spanned by the initial interaction strength $K_+(0)$ and backscattering strength $\tilde D_+(0)$. Throughout the calculation, we have set $a(0)=5~{\rm nm}$ and $v_F=1\times10^5~{\rm m/s}$, respectively.
    The other parameter values are the same as those in Fig.~\ref{fig:PD_transport1}(b).
    See Table~\ref{Table:Parameters} for the complete set of the adopted parameter values. 
    }
    \label{fig:localization_length_temperature}
\end{figure}

In conclusion, we demonstrate that nonlocal superconductivity can be fragile in the presence of pairing asymmetry, a realistic factor omitted in the previous studies~\cite{Klinovaja:2014helical,Hsu:2021}. 
As displayed in Fig.~\ref{fig:PD_transport2},
reducing $  \Delta_c(0)$ or increasing $K_+(0)$ quickly shifts the balance toward local pairing and expands the asymmetric and mixed superconducting regions. We also note an interesting effect of asymmetry on backscattering: regions that are fully insulating in the symmetric case can transition into phases where only one channel localizes while the other remains a helical liquid. Thus, pairing imbalance can lift the system out of the fully insulating regime and restore metallic behavior in one of the channels.

\subsection{Localization length and temperature in the insulating phase}
\label{subsec:localization}

In the fully insulating phase, the helical channels become effectively gapped, and electrical transport is suppressed once the channel length exceeds the localization length $\xi_{\text{loc}}$ and the temperature falls below the localization scale $T_{\text{loc}}$. In the RG description, this regime corresponds to both $\tilde D_1$ and $\tilde D_2$ flowing to strong coupling, from which $\xi_{\text{loc}}$ and $T_{\text{loc}}$ can be estimated. 
To this end, we observe that the RG flow is terminated at the scale $l^*$, which corresponds to a physical length, 
\begin{equation}
    \xi_{\text{loc}} = a (l^*) \approx a (0) e^{l^*}.
    \label{Eq:xi_loc}
\end{equation}  
The corresponding localization temperature can be computed as $  T_{\rm loc}=  \hbar u_n(l^*) / (k_B \xi_{\rm loc}) $, where $u_n(l^*)$ is the smaller of the two renormalized velocities, serving as an effective velocity characterizing the onset of localization, as both channels become localized near $l^*$.

Using the numerical values of ${l^*}$,
we compute $\xi_{\text{loc}}$ in Eq.~\eqref{Eq:xi_loc} and $T_{\text{loc}}$, going beyond typical estimation from the scaling dimension of Eq.~\eqref{eq:rs_action}. 
The results are summarized in 
Fig.~\ref{fig:localization_length_temperature}. Across the insulating region, the localization length varies with both the interaction strength and the magnitude of spin-flip backscattering. Stronger repulsive interactions or larger initial backscattering lead to a smaller $l^{*}$ and therefore a shorter localization length, which corresponds to a higher localization temperature $T_{\rm loc}$. Additionally, the estimation shows a weaker tendency towards localization than the corresponding scales in typical nonhelical channels subjected to comparable disorder strength, consistent with the enhanced robustness of helical edge transport~\cite{Altshuler:2013}.

In addition to transport properties, the RG flow of the couplings enables analysis of the topological characteristics, which we explore below.

\section{Topological phase diagrams} 
\label{sec:topological_properties}

In this section, we first derive the formula for the number of Majorana zero modes in Sec.~\ref{subsec:Majorana_criterion} from an effective model.
Combining with the RG results in Sec.~\ref{subsec:topo_PD}, we obtain the corresponding topological phase diagrams under the interplay between the Coulomb interaction, superconductivity and spin-flip backscattering.

\subsection{Characterizing topological phases}
\label{subsec:Majorana_criterion}

The above RG analysis of transport identifies two limiting regimes in which the topological behavior can be inferred directly from the relevance of random spin-flip backscattering. When the backscattering is relevant, the system flows toward an insulating phase, which is bulk localized and topologically trivial. In the opposite limit, where the backscattering flows to zero, the clean limit considered in earlier works~\cite{Klinovaja:2014helical,Hsu:2018} is recovered.

More importantly, an intermediate regime exists in which the backscattering strength remains finite at the RG stopping scale. 
In this regime, spin-flip backscattering is expected to influence the topology and cannot be neglected. Although the random backscattering amplitude vanishes upon disorder averaging at the Hamiltonian level, its variance remains finite and produces nonzero physical effects, as indicated by the effective action. To capture this influence within an effective description, we replace the random term by a uniform backscattering term whose amplitude is set by the root-mean-square strength extracted from the RG flow, and construct the corresponding effective Bogoliubov-de Gennes Hamiltonian. This procedure allows us to determine the topological phase boundaries and the associated number of Majorana zero modes. The root-mean-square amplitude is not a unique choice for representing the disorder scale, but it provides a natural measure of the finite backscattering fluctuations that remain after disorder averaging.

The resulting low-energy regimes can be characterized by the number of Majorana zero modes localized near the ends of the superconducting region, which coincide with the system corners in the geometry considered here.\footnote{In the geometry shown in Fig.~\ref{fig:setup}(a), Majorana zero modes appear near the physical corners of the device. The term ``corner'' therefore refers only to the device geometry and does not necessarily imply higher-order topology.}
To determine this number, we construct an effective fermionic description from the renormalized coupling strengths determined from the above RG analysis. This procedure is motivated by refermionization: at the refermionizable points, the competing cosine operators in the bosonized Hamiltonian are mapped onto fermionic bilinears, with the superconducting and spin-flip backscattering couplings acting as competing mass terms. Although exact refermionization applies only at particular values of the interaction parameters, the renormalized regimes considered here can be adiabatically continued to the corresponding refermionizable or noninteracting limits without closing the bulk gap~\cite{Gangadharaiah:2011,Hsu:2018,Thakurathi:2018}; see also Appendix~\ref{Appendix:refermionization} for the details. Moreover, this continuation preserves the hierarchy among the relevant coupling strengths, and thus the topological properties remain unchanged.

We remark that the resulting Bogoliubov-de Gennes Hamiltonian should not be interpreted as a return to the bare noninteracting problem. Its parameters are the coupling strengths obtained from the RG flow and therefore encode the effects of interactions accumulated down to the RG stopping scale. The effective Hamiltonian is used to determine the competition among these renormalized mass terms and to identify the associated bulk-gap closings and number of Majorana zero modes. Its role is thus to diagnose the topology of the RG low-energy theory, rather than to reproduce the full interacting dynamics.

We accordingly introduce the effective single-particle description, $H_{\rm eff} = \frac{1}{2} \sum_{k} \Psi_{k}^\dagger  \mathcal{H}_{\rm eff} (k) \Psi_{k} $, 
with the Nambu spinor, 
\begin{subequations}
\begin{equation}
    \Psi_k = (R_{1\downarrow},L_{1\uparrow},R_{2\downarrow},L_{2\uparrow},R_{1\downarrow}^\dagger,L_{1\uparrow}^\dagger,R_{2\downarrow}^\dagger,L_{2\uparrow}^\dagger)^T,
\end{equation}
and the Hamiltonian density, 
\begin{align}
    \mathcal{H}_{\rm eff} (k) =& 
    \hbar v_{F} k 
    \sigma^z
    - \Delta_c(l^{*}) \eta^y \tau^x  \sigma^y \nonumber 
    \\ 
    &  - \Delta_+(l^{*}) \eta^y  
    \sigma^y
    - \Delta_-(l^{*}) \eta^y  \tau^z \sigma^y \nonumber  \\
   &  
    + V_+(l^{*}) \eta^z  
    \sigma^x 
    + V_-(l^{*}) \eta^z  \tau^z  \sigma^x , 
\end{align}
\end{subequations}
with Pauli matrices $ \eta^{\mu}$, $ \tau^{\mu}$, and $ \sigma^{\mu}$ acting on the particle-hole, channel, and spin subspaces, respectively, with $\mu \in \{ 0, x, y, z \}$. In this formulation, the parameters are given by their renormalized values obtained from the RG flow,  
including the effective root-mean-square  amplitudes, $V_{\pm}(l^*)\equiv [V_1(l^*) \pm V_2(l^*) ]/2$, of the random spin-flip strengths defined in Eq.~\eqref{rs_hamiltonian}.
The presence of spin-flip terms leads to time-reversal symmetry breaking and the system now belongs to class BDI; see Appendix~\ref{Appendix:single-particle-analysis} for details. 
Since the topology is characterized by a $\mathbb {Z}$ invariant,  we  expect additional topological phase transitions upon varying the system parameters.

With the effective model, we solve the corresponding Bogoliubov-de Gennes equation and find the Majorana zero-energy solutions in the parameter space; see the algebra details in Appendix~\ref{Appendix:single-particle-analysis}.
We derive the analytic expression of the number of zero modes as a function of the system parameters. Belonging to the BDI class, the topological phases are  characterized by a $\mathbb Z$-valued 
invariant, 
\begin{widetext}
    \begin{align}
\label{eq:Nmzm} 
    N_{\text{mzm}} =& \sum_{\varepsilon \in \{+,-\}} \Theta \left( - |\Delta_-(l^{*}) + \varepsilon V_{-}(l^{*})| + |\Delta_+(l^{*}) + \varepsilon V_{ +}(l^{*}) | \right)   \nonumber \\
&\hspace{80pt} \times \Theta (  \sqrt{ [\Delta_-(l^{*}) + \varepsilon V_{ -}(l^{*}) ]^2 + [ \Delta_c(l^{*}) ]^2} - |\Delta_+(l^{*}) + \varepsilon V_{ +}(l^{*}) | ),
\end{align} 
\end{widetext}
with the Heaviside step function $\Theta(x)$; $\Theta(0)$ is immaterial here. The analytical formula in Eq.~\eqref{eq:Nmzm} is one of the main findings in this work, which also serves as the guidance for the topological phase diagrams presented below.

Interestingly, a given $\varepsilon$ branch contributes one zero mode (per system corner) when the following inequality is satisfied,
\begin{align}
\label{eq:Nmzm_equality}
|\Delta_-(l^{*}) + \varepsilon V_-(l^{*})| < & |\Delta_+(l^{*}) + \varepsilon V_+(l^{*})| \nonumber \\
< & \sqrt{[ \Delta_-(l^{*}) + \varepsilon V_-(l^{*}) ]^2 + [\Delta_c(l^{*})]^2} .
\end{align}
Thus, a zero mode in branch $\varepsilon$ emerges only if the quantity $|\Delta_+(l^{*}) + \varepsilon V_+(l^{*})|$ lies between the lower and upper bounds, which can be adjusted through the pairing asymmetry. 

The above relation also clarifies how backscattering modifies the topological invariant. 
Namely, in the absence of backscattering, $V_\pm(l^{*}) =0$, both branches share identical bounds, so that the two Majorana zero modes appear or disappear simultaneously, recovering the $\mathbb{Z}_2$ topological invariant reported in previous works~\cite{Klinovaja:2014helical,Hsu:2018}.
The inclusion of backscattering breaks this locking by inducing opposite shifts $\pm V_\pm (l^{*})$ in the two $\varepsilon$ branches. The conditions for the two $\varepsilon$ branches therefore become detuned, giving rise to a new topological regime with $N_{\mathrm{mzm}} = 1$. As a consequence, the topological invariant is characterized by $\mathbb{Z}$, consistent with the symmetry class BDI in the presence of the spin-flip terms.

\begin{figure*}[t]
\centering
\includegraphics[width=1.0\linewidth]{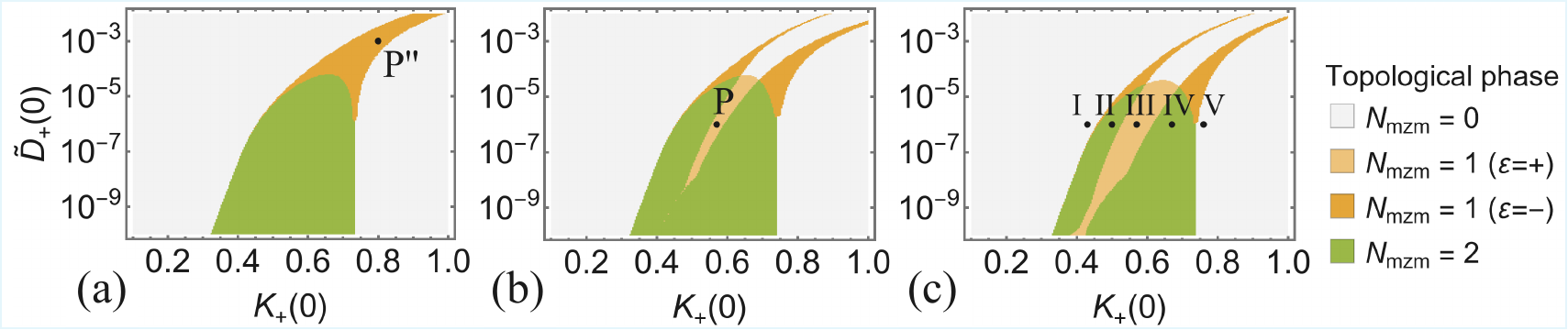}
    \caption{Topological phase diagrams in the $[K_+(0),\,\tilde D_+(0)]$ plane, characterized by the zero mode number $N_{\rm mzm}$ per system corner obtained from the renormalized couplings and Eq.~\eqref{eq:Nmzm}. We further distinguish the $N_{\rm mzm}=1$ phase according to which of the $\varepsilon$ conditions in Eq.~\eqref{eq:Nmzm_equality} is fulfilled.
    The parameters include (a) $\tilde \Delta_-(0) = K_-(0) =0$, (b) the pairing asymmetry case with $\tilde \Delta_-(0)>0$ and (c) the Coulomb asymmetry case with $K_-(0)>0$.     
    See Table~\ref{Table:Parameters} for the complete set of the adopted parameter values.  
    }
    \label{fig:PD1_Nmzm}
\end{figure*}

As a result, spin-flip backscatterings provide a mechanism to detune the two  zero-mode conditions, thereby generating a topological region in which only one of them is satisfied. From an experimental perspective, this implies that moderate disorder can introduce a practical tuning knob rather than merely a detrimental perturbation. Because the $N_{\mathrm{mzm}} = 1$ region occupies a finite region in parameter space, the resulting single-zero-mode phase can still exist against sufficiently small variations in pairing strengths, interaction asymmetries, and disorder amplitude.    

Another important implication of our finding is related to the $\pi$-junction setups~\cite{Schrade:2015,Laubscher:2020}, in which the signs of the induced local pairing are opposite in the two channels. In our notation, this corresponds to $\Delta_+=0$ and $\Delta_-\ne 0$; the pure $\pi$-junction additionally satisfies $\Delta_c=0$, but keeping $\Delta_c$ explicit is useful for assessing its effect on the zero modes. 
With this setting, Eq.~\eqref{eq:Nmzm_equality} takes the form
\begin{align}
\label{eq:Nmzm_equality_pi}
|\Delta_-(l^{*}) + \varepsilon V_-(l^{*})| < & | V_+(l^{*})| \nonumber \\
< & \sqrt{[ \Delta_-(l^{*}) + \varepsilon V_-(l^{*}) ]^2 + [\Delta_c(l^{*})]^2} .
\end{align}
It is instructive to compare our findings with previous studies on time-reversal invariant topological superconductors in bilayer systems, such as Ref.~\cite{Laubscher:2020}. In those setups, 
the driving mechanism for Majorana zero modes is to have a dominant coherent single-particle interlayer tunneling over the pairing~\cite{Laubscher:2020}. 
In contrast, the driving mechanism in Eq.~\eqref{eq:Nmzm_equality_pi} is the nonlocal pairing $\Delta_c$. When we neither have this term nor intelayer tunneling, the clean limit of a $\pi$-junction  ($V_{\pm}=0,\,\Delta_c=0$) would result in the impossible condition $|\Delta_-(l^{*})| <  0$ in Eq.~\eqref{eq:Nmzm_equality_pi}.
Thus,  a pure $\pi$-phase difference is insufficient to stabilize Majorana zero modes. 

However, the inclusion of spin-flip backscattering and nonlocal pairing qualitatively alters the behavior of the topological phase transition. As indicated by Eq.~\eqref{eq:Nmzm_equality_pi}, sufficiently strong backscattering allows the lower bound of the criterion to be satisfied, while a finite nonlocal pairing $\Delta_{c}(l^{*})$ opens a finite parameter regime that supports Majorana zero modes. The resulting topological phase therefore emerges from a cooperative mechanism involving magnetic disorder and nonlocal pairing.
These observations highlight that spin-flip backscattering introduces additional parameter windows for topological phases rather than simply suppressing them. 

As a complementary viewpoint to the refermionization analysis above, one can consider the spin-difference parity in the clean limit. In clean double helical channels, one can show that the parity associated with the spin difference between the two channels is conserved~\cite{HsuPhonons:2024}. When the nonlocal pairing is dominant, this parity protects the corresponding ground-state degeneracy, providing an alternative way to identify the nontrivial topology without invoking refermionization.
When random spin-flip backscattering is present but RG irrelevant, its coupling flows to zero at low energies. The system then approaches the same low-energy fixed point as the clean case, so the parity-based argument remains applicable. Therefore, the topological phase can survive over finite regions of the parameter space.

By contrast, when random spin-flip backscattering is RG relevant, it drives the helical channels toward an insulating phase, which is topologically trivial. 
When the backscattering strength instead flows to a small but finite value, the parity argument is no longer sufficient. We then apply the analytical criterion developed above together with the RG flow to determine the resulting topological phase. In the following section, we use this combined framework to construct phase diagrams for the interacting double helical channels.

\subsection{Topological phase diagram}
\label{subsec:topo_PD}

Here we use the renormalized couplings for given sets of initial parameters, as demonstrated in Sec.~\ref{sec:transport}, to compute $N_{\mathrm{mzm}}$ using Eq.~\eqref{eq:Nmzm}. 

\subsubsection{Varying interaction and backscattering strengths}
\label{subsubsec:topo_Kplus_Dplus}

 In Fig.~\ref{fig:PD1_Nmzm}, we  present the phase diagrams based on the analytically obtained $N_{\mathrm{mzm}}$. An immediate observation is the clear correlation between the topological character and the nature of the superconducting phases. The regions hosting Majorana zero modes ($N_{\rm mzm} > 0$) tend to overlap with those where nonlocal pairing $\tilde \Delta_c$ dominates under RG flow. This confirms that interaction-driven enhancement of nonlocal pairing is the key mechanism for realizing topological phases in this system, consistent with previous findings~\cite{Hsu:2018}. 

In the fully symmetric case shown in Fig.~\ref{fig:PD1_Nmzm}(a), a robust topological region with $N_{\rm mzm}=2$ appears for intermediate repulsive interactions and weak backscattering strengths. For a fixed interaction strength within this region, increasing $\tilde D_+(0)$ can induce a sequence of topological phase transitions from $N_{\text{mzm}}=2$ to $N_{\text{mzm}}=1$, and finally to the trivial phase $N_{\text{mzm}}=0$. This demonstrates that the Majorana modes are sensitive to spin-flip backscattering, which acts to close the topological gap.

\begin{figure*}[ht]
\centering
\includegraphics[width=1.0\linewidth]{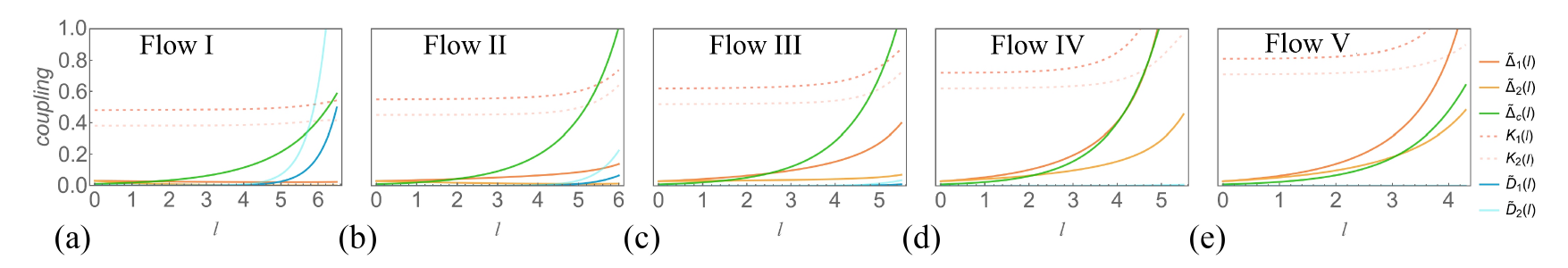}
    \caption{RG flows  I--V, with the adopted parameter values indicated in Fig.~\ref{fig:PD1_Nmzm}(c). See Table~\ref{Table:Parameters} for the complete set of the adopted parameter values. 
    }
    \label{fig:flow_I_to_V}
\end{figure*}

Going beyond the symmetric channel setting, we introduce pairing asymmetry in Fig.~\ref{fig:PD1_Nmzm}(b) while keeping all the other parameters identical to Fig.~\ref{fig:PD1_Nmzm}(a). Most of the $N_{\rm mzm}=2$ region  survives, but is split by a narrow, wedge-shaped region in which the system instead hosts a single Majorana zero mode. 
A representative point P inside this wedge is marked in the figure, corresponding to the same parameter set as the dot in Fig.~\ref{fig:PD_transport1}(b). 
The origin of this wedge can be understood directly from  Eq.~\eqref{eq:Nmzm_equality}. Namely, since the renormalized backscattering asymmetry $V_{-}(l^{*})$ remains negligible with $\tilde D_n$ and $K_n$ flowing symmetrically, we have  
\begin{equation}
\label{eq:inequality_symm_dis}
    |\Delta_-(l^{*})| < |\Delta_+(l^{*}) + \varepsilon V_{ +}(l^{*})| < \sqrt{ [\Delta_-(l^{*})]^2 + [\Delta_c(l^{*}) ]^2}. 
\end{equation}
Unlike the symmetric pairing case, where the lower bound is zero, a finite $\tilde{\Delta}_-(0)$ here shifts the lower bound to a nonzero value. As a result, the condition for the $\varepsilon = -$ branch is not satisfied in a certain region near point P. 
This behavior indicates that one of the Majorana zero mode conditions tends to be more sensitive to pairing asymmetry.

We now look into the effects of Coulomb asymmetry and present Fig.~\ref{fig:PD1_Nmzm}(c).
Even though the initial local pairings are symmetric, the RG flow generates finite $\tilde \Delta_-(l^{*})$ and $\tilde D_-(l^{*})$ due to the difference in $K_{n}$. Similarly to Fig.~\ref{fig:PD1_Nmzm}(b), the imbalance between the local pairings leads to disappearance of one of the Majorana zero modes in the otherwise uniform $N_{\rm mzm} = 2$ region. 
 
Interestingly, as seen in Figs.~\ref{fig:PD1_Nmzm}(b,c), an asymmetry in the pairing strengths or Coulomb interactions can lead to an {\it electrically tunable revival of Majorana zero modes}.
Namely, upon varying the interaction parameters, the Majorana zero mode associated with the $\varepsilon = -$ condition first disappears and then reemerges.
On the other hand, the $\varepsilon = +$ mode remains stable over a broader range for weak disorder strengths, or can be destabilized by a strong disorder. 
To further examine this feature, we select the parameter sets I--V [marked in Figs.~\ref{fig:PD1_Nmzm}(c)] and present their RG trajectories in Fig.~\ref{fig:flow_I_to_V}.
The plots illustrate how the renormalized couplings evolve as the system passes through the splitting of the $N_{\mathrm{mzm}} = 2$ region. From I to V, the dominating couplings change as $K_+(0)$ is continuously increased. Specifically, backscattering dominates in Flow~I, leading to a trivial insulating phase. For Flow~II, the nonlocal pairing $\tilde{\Delta}_c$ becomes the sole relevant coupling, leading a nontrivial phase with $N_{\mathrm{mzm}} = 2$. Flows~III and IV show the onset of competition between the local pairings and $\tilde{\Delta}_c$, consistent with the change in $N_{\mathrm{mzm}} $, while in Flow~V, $\tilde{\Delta}_1$ eventually dominates, driving the system back to a trivial phase.
The topological phase diagrams in Fig.~\ref{fig:PD1_Nmzm} are organized by varying the interaction and backscattering strengths; for completeness, we present additional analysis of the pairing asymmetry effects at a fixed interaction strength in Fig.~\ref{fig:PD2_Nmzm} in Appendix~\ref{Appendix:more}.

Intriguingly, our findings suggest the emergence of a cascade of successive topological phase transitions, driven by tuning the interaction strength within the helical channels. Here, the cascade is manifested by stepwise changes in the number of Majorana zero modes across multiple bulk-gap closings as the renormalized couplings cross distinct topological phase boundaries. Such interaction tuning can be experimentally realized through screening control~\cite{HsuPhonons:2024}.

In addition to  the series of topological phase transitions and the phenomenon of Majorana revival, our results also reveal imperfection-induced topological phases, which we elaborate next.

\subsubsection{Imperfection-induced topological phases}
\label{subsubsec:topo_Nmzm1}

In this section, we discuss an interesting feature in Fig.~\ref{fig:PD1_Nmzm}(a) appearing at weak interactions. Here, increasing backscattering strength while fixing all the other initial parameter values can move the system from the trivial region into $N_{\rm mzm}=1$.
To understand this rather counterintuitive behavior, we take a closer look at a representative point in the $N_{\rm mzm}=1$ region, marked as P$^{\prime\prime}$ in Fig.~\ref{fig:PD1_Nmzm}(a).
Despite the exact symmetry of the two channels under RG (see the corresponding flow in Fig.~\ref{fig:flow_Ppp} in Appendix~\ref{Appendix:more_on_Nmzm1}),
the renormalized backscattering strength shifts the combinations $|\Delta_+(l^{*}) \pm V_{ +}(l^{*})|$ in Eq.~\eqref{eq:inequality_symm_dis} in opposite directions.
At the cutoff $l^*$, one of these combinations satisfies the inequality while the other does not, placing the system in the single zero mode regime. 
Notably, this phenomenon occurs over a rather broad region of parameter space. 
As shown in Fig.~\ref{fig:backscattering_induced_Nmzm1} for even weaker interaction strengths (see Appendix~\ref{Appendix:more_on_Nmzm1}), we observe that, for sufficiently large pairing asymmetry, 
increasing the backscattering strength drives the system into the $N_{\mathrm{mzm}} = 1$ phase from a topologically trivial phase in the clean limit. 
In conclusion, the interplay among interactions, disorder-induced backscattering, and pairing asymmetry can induce the emergence of Majorana zero modes from a topologically trivial phase in the clean limit.

In the following section, we explore a broader parameter space by constructing three-dimensional phase diagrams that capture the interplay among interactions, disorder effects, and proximity-induced pairing.

\section{Phase diagrams in three-dimensional parameter space}
\label{sec:3dphasediagrams}

In this section, we present three-dimensional phase diagrams that elucidate the intertwined roles of interactions, disorder-induced backscattering, and various pairing processes in helical channels. As above, the phase structure is analyzed from both transport and topological perspectives for complementary representations.

In the first set of phase diagrams, we vary the average interaction strength $K_{+}(0)$, the average local pairing strength $\tilde{\Delta}{_+}(0)$, and the average backscattering strength $\tilde{D}{_+}(0)$, thereby highlighting the competition between local and nonlocal pairing processes in the presence of interactions and disorder. In the second set, $\tilde{\Delta}{_+}(0)$ is replaced by the pairing asymmetry $\tilde{\Delta}_{-}(0)$, allowing us to directly isolate and visualize the impact of pairing imbalance on the phase structure.
This formulation provides a unified view of how interactions, disorder, and pairing asymmetry cooperate to shape transport behavior and topological stability.

\begin{figure*}[t]
    \centering
    \includegraphics[width=1.0\linewidth]{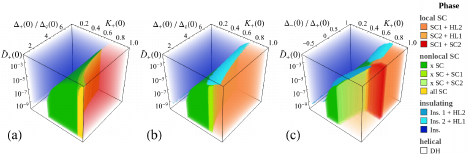}
    \caption{Transport-based phase diagrams in three-dimensional parameter space, including (a--b) local-to-nonlocal pairing ratio $\Delta_+(0)/\Delta_c(0)$ 
    and (c)  pairing asymmetry $\tilde{\Delta}{_-}(0)$. 
    The regions corresponding to the insulating phase (blue) and the local superconducting phase in channel 2 (orange) or channels 1 and 2 (red) are rendered with increased transparency to reveal the internal structure.
The adopted parameter values include (a) $\tilde \Delta_-(0) = 0$, (b) $\tilde \Delta_-(0) = 0.01$, and (c) $\tilde \Delta_+(0)=0.03$. In each case, we set $\tilde \Delta_c(0)=0.01$, and $K_-(0)=\tilde D_-(0)=0$.
See Table~\ref{Table:Parameters} for the complete set of the adopted parameter values.
}
    \label{fig:PD_3D}
\end{figure*}

We first discuss the phase diagrams with a varying average local pairing strength, as shown in Fig.~\ref{fig:PD_3D}(a--b). We observe that increasing the local pairing strength suppresses the nonlocal superconducting phase in both symmetric and asymmetric cases, as displayed in Fig.~\ref{fig:PD_3D}(a) and Fig.~\ref{fig:PD_3D}(b), respectively. Additionally, the asymmetry introduces distinct features in Fig.~\ref{fig:PD_3D}(b). Specifically, we observe the emergence of the asymmetric local superconducting phase at weak interactions and an asymmetric insulating phase (light blue), where one channel localizes faster than the other. 

\begin{figure*}[t]
    \centering
    \includegraphics[width=1.0\linewidth]{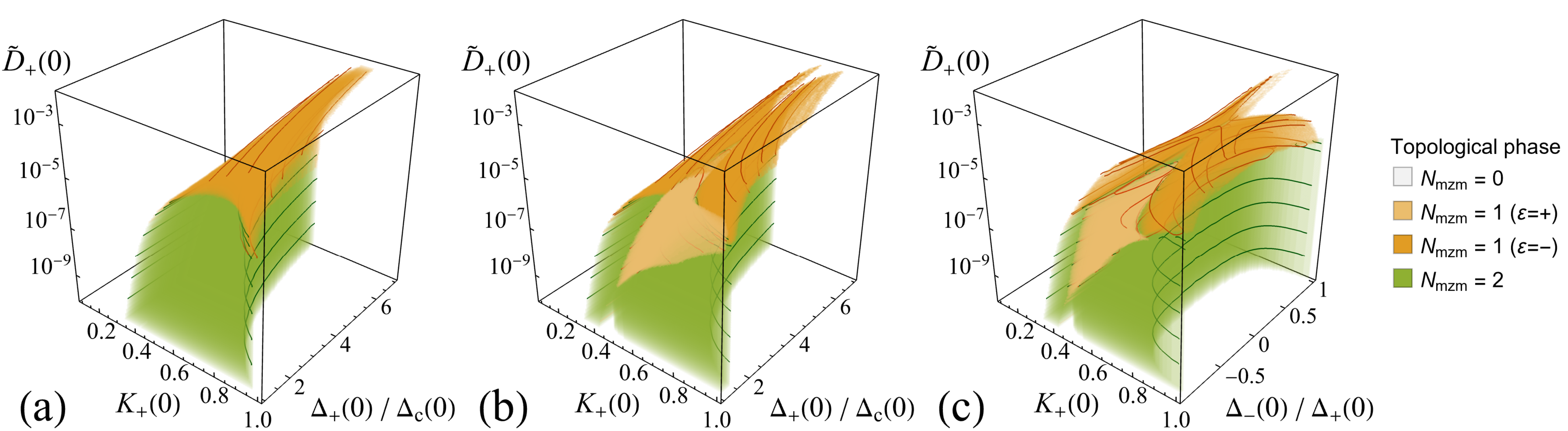}
    \caption{Topological phase diagrams  in three-dimensional parameter space, including (a--b) local-to-nonlocal pairing ratio $\Delta_+(0)/\Delta_c(0)$ 
    and (c)  pairing asymmetry $\tilde{\Delta}{_-}(0)$. 
The adopted parameter values include (a) $\tilde \Delta_-(0) = 0$, (b) $\tilde \Delta_-(0) = 0.01$, and (c) $\tilde \Delta_+(0)=0.03$. In each case, $\tilde \Delta_c(0)=0.01$, and $K_-(0)=\tilde D_-(0)=0$.
Solid colored curves indicate representative phase boundaries at fixed $\tilde D_+(0)$ planes, illustrating the evolution of the boundaries for several representative values of the backscattering strength.
See Table~\ref{Table:Parameters} for the complete set of the adopted parameter values. 
}
\label{fig:PD_3D_Nmzm}
\end{figure*}

The corresponding topological phase diagrams in Fig.~\ref{fig:PD_3D_Nmzm}(a--b) confirm the robustness of the $N_{\rm mzm}=2$ phase at a weak backscattering when the nonlocal pairing dominates. In the symmetric case in Fig.~\ref{fig:PD_3D_Nmzm}(a), increasing  backscattering strength eventually destroys the Majorana zero modes. Interestingly, the three-dimensional phase diagrams reveal that the $N_{\mathrm{mzm}} = 1$ phase emerges over a broad region of parameter space. As illustrated in the asymmetric case in Fig.~\ref{fig:PD_3D_Nmzm}(b), a robust $N_{\mathrm{mzm}} = 1$ region appears between the $N_{\mathrm{mzm}} = 2$ phases and remains accessible in a wide range of backscattering strengths. 
As the local pairing strength increases, this intermediate phase eventually shrinks. The three-dimensional plots clearly reveal that multiple phase transitions and the revival of zero modes can be induced by tuning a single control parameter, such as the backscattering strength (via external magnetic fields), the interaction strength (through screening effects), or the balance between local and nonlocal pairing (controlled by channel separation or the proximity interface), as well as the combinations of multiple parameters. 

Next, we discuss the transport-based phase diagram in Fig.~\ref{fig:PD_3D}(c), which incorporates pairing asymmetry.  
Consistent with previous observations, we note that increasing pairing asymmetry shifts the system toward regions where asymmetric phases are favored, such as the asymmetric local superconducting phases at weak interactions, and in some regions, asymmetric insulating phases, where one of the channels localizes faster.

The corresponding topological phase diagram with pairing asymmetry is shown in Fig.~\ref{fig:PD_3D_Nmzm}(c), complementing Figs.~\ref{fig:PD_3D_Nmzm}(a--b). Consistent with the analysis in the previous section, the $N_{\mathrm{mzm}} = 1$ phase emerges immediately upon deviating from the perfectly symmetric limit $\tilde{\Delta}_-(0) = 0$ and remains accessible over a broad range of backscattering strengths. This has important experimental implications, as pairing imbalance between channels and weak random spin-flip backscattering are ubiquitous in realistic systems.

Finally, we briefly remark on our findings in view of the $\pi$-junction limit discussed in Sec.~\ref{sec:topological_properties}. An ideal $\pi$-junction corresponds to the limit in which the local pairings in the two channels have opposite signs, thus $|\Delta_{-}(0)/\Delta_{+}(0) | \to  \infty$. Our analysis shows that the presence of a residual uniform component $\Delta_{+}$ can stabilize multiple phases, revealing rich behavior in nonideal $\pi$ junctions. Moreover, the persistence of the topological phase in Fig.~\ref{fig:PD_3D_Nmzm}(c) in the presence of disorder indicates that $\pi$-junction-based setups can have robust topological phases against realistic backscattering effects.

\section{Microscopic features of the Majorana zero modes} 
\label{sec:microscopic_features}

\begin{figure}[t]
    \centering
    \includegraphics[width=1.0\linewidth]{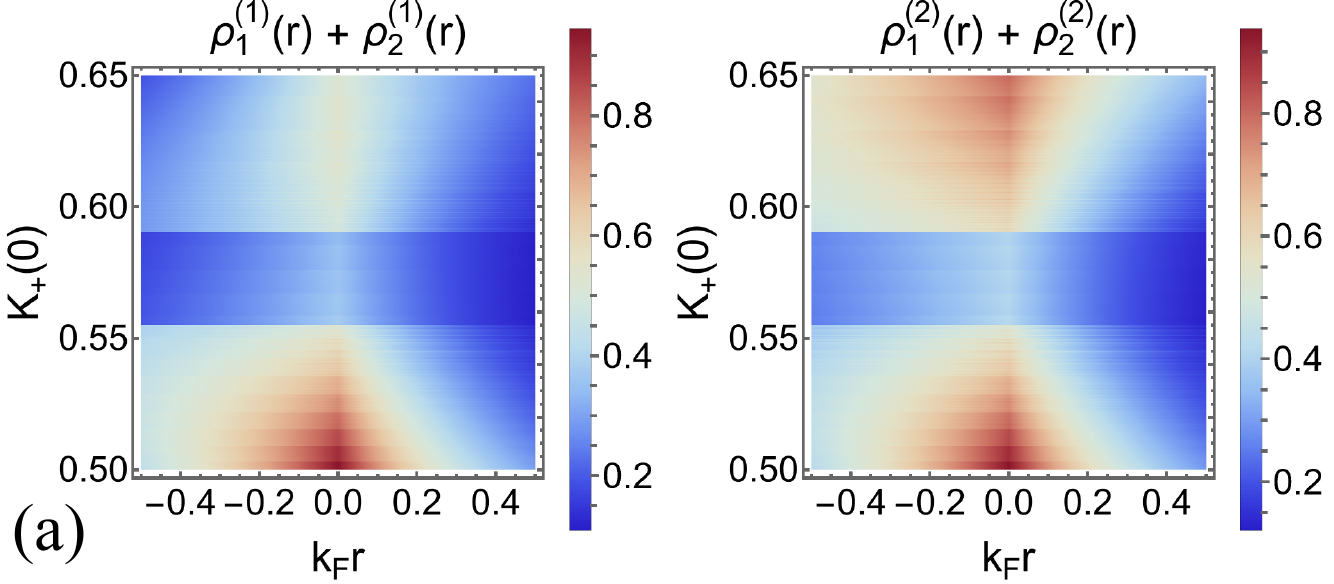}
    \includegraphics[width=1.0\linewidth]{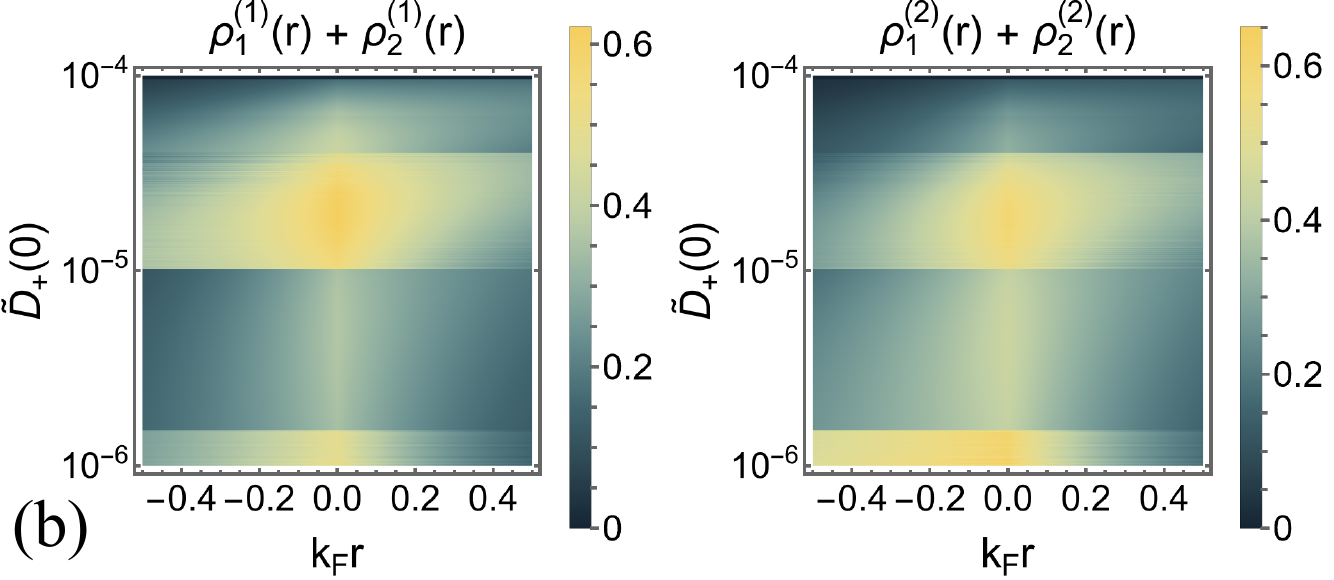}
    \caption{Channel-resolved total density profiles $\sum_j \rho_j^{(n)}(r)$ for channel $n=1$ (left panels) and $n=2$ (right panels). (a) Density evolution as a function of $K_+(0)$ with fixed $\tilde D_+(0)=10^{-6}$. (b) Density evolution as a function of $\tilde D_+(0)$ with fixed $K_+(0)=0.6$. The remaining fixed parameters correspond to point P in Fig.~\ref{fig:PD1_Nmzm}(b), $\tilde \Delta_+(0)=0.03$, $\tilde \Delta_-(0)=0.01$, $\tilde \Delta_c(0)=0.01$, and $\tilde D_-(0)=K_-(0) =0$. See Table~\ref{Table:Parameters} for the complete set of the adopted parameter values.}
    \label{fig:K_D_densityplot}
\end{figure}

In this section, we examine how the Majorana zero modes evolve through the series of topological phase transitions and discuss their features in local probes.
To this end, we focus on the representative parameter sets near the point P marked in Fig.~\ref{fig:PD1_Nmzm}(b). In this region, tuning either the electron-electron interaction $K_+(0)$ or the backscattering $\tilde D_+(0)$ strength alone results in the phase transitions with the sequence $N_{\rm mzm}=2\rightarrow1\rightarrow2$. The scanned ranges of the $K_+(0)$ and $\tilde D_+(0)$ values used in Figs.~\ref{fig:K_D_densityplot}(a--b) were chosen to illustrate this path, where one of the Majorana zero modes is destroyed and then reappears, signifying a tunable revival.

To visualize how the modes evolve, we compute their corresponding  wavefunctions following the algebra in Appendix~\ref{Appendix:single-particle-analysis}, with the renormalized couplings acquired from the above RG analysis. 
After obtaining the zero-energy orthonormal modes, $\Phi_{\rm mzm,1}(r)$ and $\Phi_{\rm mzm,2}(r)$, we reconstruct the full wavefunctions by reinstalling the fast-oscillating factors $e^{\pm i k_F r}$, 
\begin{equation}
    \Psi_{\text{mzm},j}(r) = e^{i k_F r  \eta^z \sigma^z} \Phi_{\text{mzm},j}(r),
\end{equation}
expressed in the physical basis $(\psi_{1\downarrow},\psi_{1\uparrow},\psi_{2\downarrow},\psi_{2\uparrow},\psi_{1\downarrow}^\dagger,\psi_{1\uparrow}^\dagger,\psi_{2\downarrow}^\dagger,\psi_{2\uparrow}^\dagger)^T$. 
To proceed, we define 
\begin{equation}
    \rho_j(r)\equiv |\Psi_{\text{mzm},j}(r)|^2.
\end{equation}
for the $j$th zero mode, with $j\in\{1,2\}$. 
Since the Majorana wavefunctions $\Psi_{\text{mzm},j}(r)$ satisfy self-conjugation, the electron and hole components contribute equally. Consequently, $\rho_j(r)$ corresponds to twice the particle component of the density profiles. 
Notably, the wavefunction has finite support in both channels, reflecting the composite nature of the zero modes discussed in Appendix~\ref{Appendix:single-particle-analysis}.

Motivated by the local scanning probes with high spatial resolution, we examine the channel-resolved density profiles. To this end, we  decompose $\rho_j(r)$ into  $\rho_j^{(n)}(r)\equiv |\Psi^{(n)}_{\text{mzm},j}(r)|^2$ by projecting
\begin{eqnarray}
    \Psi_{\text{mzm},j}^{(1,2)}(r) \equiv \left[ \eta^0 \left( \frac{\tau^0\pm\tau^z}2 \right) \sigma^0 \right] \Psi_{\text{mzm},j}(r), 
\end{eqnarray}
and evaluate them using the RG results. 

To visualize the evolution of the zero modes accross the topological phase transitions, Figs.~\ref{fig:K_D_densityplot}(a,b) display the spatial density maps as functions of the interaction parameters $K_+(0)$ and backscattering strengths $\tilde D_+(0)$. Here, we plot the total density in each channel, $\sum_{j}\rho_j^{(n)}(r)$, combining the density of all the zero modes present in the system. 
In contrast to nonhelical systems~\cite{Klinovaja:2012a}, the resulting density profiles here do not exhibit spatial oscillations, reflecting the underlying helical nature of the channels.

We first discuss Fig.~\ref{fig:K_D_densityplot}(a), which tracks the density evolution as the interaction parameter $K_+(0)$ is varied. Throughout this range, the renormalized nonlocal pairing $\tilde \Delta_c (l^*)$ remains the dominant scale, while the local pairings $\tilde \Delta_{1,2}(l^*)$ gradually increase and the backscattering strengths $\tilde D_{1,2}(l^*)$ gradually decrease as $K_+(0)$ is raised.
At stronger interactions, the topological criterion in Eq.~\eqref{eq:Nmzm_equality} is satisfied for both $\varepsilon=+$ and $\varepsilon=-$, and the system hosts two well-localized zero modes around $r=0$. On the $r<0$ side, where the nonlocal pairing is absent, the modes appear more extended. This follows directly from the form of the decay constants on that side: as given in Eq.~\eqref{eq:kappa_less}, $\kappa_{n,\varepsilon}^{<} = \big|\Delta_n + \varepsilon V_{n}\big|/(\hbar v_F)$, which depends only on the local channel parameters. By contrast, on the $r>0$ side the decay constants $\kappa_{\lambda,\varepsilon}^{>} = \left|(\Delta_+ + \varepsilon V_+) + \lambda \sqrt{(\Delta_- + \varepsilon V_-)^2 + \Delta_c^2}\right|/(\hbar v_F)$ are enhanced by the presence of $\Delta_c$, leading to sharper localization. 
As the interactions weaken (middle region), only one of the two zero-mode conditions remains satisfied, leaving a single Majorana zero mode, in line with the behavior described in Sec.~\ref{subsubsec:topo_Kplus_Dplus}. 
Further weakening the interactions (upper region) restores the second zero-mode condition, leading to the reappearance of the second topological zero mode and the enhanced total density.
At the same time, the profiles broaden compared to the lower region, and are less localized. This behavior reflects the changing balance between the local and nonlocal pairings. Namely, as the interactions weaken further, the RG flow drives the backscattering strengths $\tilde D_{1,2}(l^*)$ closer to zero, so the presence of the zero modes is set almost entirely by the pairings. In this regime, the criterion for $N_{\rm mzm}$ approaches the familiar condition $\Delta_c^2>\Delta_1\Delta_2$ in Ref.~\cite{Hsu:2018}. Eventually the local pairings
become comparable to the nonlocal one, and the Majorana zero modes extend over longer distances and eventually vanish.

In Fig.~\ref{fig:K_D_densityplot}(b), we vary the backscattering strength $\tilde D_+(0)$ to trace a vertical path, with $N_{\rm mzm}=2\to 1\;(\varepsilon=+)\to2 \to 1\;(\varepsilon=-)$ in Fig.~\ref{fig:PD1_Nmzm}(b). Deep inside the topological phase centered near $\tilde D_+(0)\sim 10^{-5}$, the zero modes exhibit their sharpest localization and highest intensity at $r=0$. However, as $\tilde D_+(0)$ approaches the phase boundaries on either side, the confinement weakens as the gap closes and the $r=0$ intensity fades.

As mentioned above, the density maps provide a qualitative indication of the bulk gap. As derived in Eqs.~\eqref{eq:kappa_less} and \eqref{eq:kappa_more}, the spatial decay constants $\kappa$ are directly proportional to the effective gaps on the respective sides of the interface. Consequently, regions of high intensity near $r=0$ signify strong confinement and a robust topological gap. Conversely, near a phase transition where $N_{\rm mzm}$ changes, the gap closes, driving the decay constant to zero, thus serving as a visual signature of the bulk gap closing.

Accompanying the zero-energy spectroscopic signatures~\cite{Machida:2019}, the spatial structure of the Majorana zero modes shown in Fig.~\ref{fig:K_D_densityplot} can be accessed using scanning tunneling microscopy. Deep inside a topological phase, sharply localized modes are expected near the system corners ($r=0,L$). In contrast, broader spatial profiles are anticipated as the system approaches phase boundaries.

Beyond the spatial density, the internal structure of these wavefunction relates to broader discussions in the existing literature. Ref.~\cite{Sticlet:2012} retained only the electronic component when defining spin polarization, focusing on quantities accessible via spin-polarized STM, while Ref.~\cite{Szumniak:2017} incorporated both electron and hole contributions in the spin polarization to account for the electron-hole overlap effects. Ref.~\cite{Awoga:2024} focused exclusively on Majorana polarization, which is intrinsically defined using both particle and hole degrees of freedom. Importantly, these works primarily analyze bulk states at finite energies rather than the topological zero modes on which we focus here.

We mention that the density profiles displayed in Fig.~\ref{fig:K_D_densityplot} correspond to contributions only from the zero modes. Experimental probes, however, may effectively integrate contributions from states away from $E=0$. As a result, a complementary analysis of finite-energy bulk states is necessary to provide additional context, such as background contrast, for experimental characterization.
Finally, distinguishing topological Majorana zero modes from trivial bound states remains an important experimental challenge. Other indicators could be useful, including Majorana polarization~\cite{Sticlet:2012,Awoga:2024}, which quantifies particle-hole overlap, as well as the sign reversal of the spin polarization of bulk states across a topological phase transition~\cite{Szumniak:2017}.

\section{Discussion} 
\label{sec:discussion}

In this work, we investigated how Coulomb interactions, proximity-induced superconductivity, and random spin-flip backscattering collectively influence the transport and topological properties of double helical liquids. 
By introducing pairing and Coulomb asymmetry between the helical channels, we uncover additional phase transitions.
By deriving the connection between the Majorana zero mode number and the renormalized couplings, we show how pairing asymmetry and spin-flip backscattering can detune the conditions, enabling an interacting, disordered system to access an $N_{\mathrm{mzm}}=1$ phase. An intriguing feature is that, for certain interaction strengths, disorder-induced backscattering can induce Majorana zero modes. 
 
From an experimental perspective, we demonstrate that random backscattering is not merely a detrimental perturbation; it also provides an additional tuning knob that can selectively activate one branch of the Majorana zero mode criterion. Moreover, we demonstrate that at moderately strong interactions, pairing asymmetry suppresses the $N_{\mathrm{mzm}} = 2$ phase, leading to additional topological phases and resulting in a cascade of transitions and a revival of a Majorana zero mode. 
These features can be electrically accessible by controlling the screened Coulomb interactions in the edge channels.
Our analysis is also relevant to phase-controlled Josephson junctions, where access to topological phases relies on phase biasing between two channels via an external magnetic flux~\cite{Li:2016,Shabani:2016,Ren:2019,Fornieri:2019}. In such devices, imperfect magnetic shielding can in practice introduce residual fields in the conduction channels, effectively generating spin-flip backscattering. Taken together, these considerations suggest that systems with controlled spin-flip backscattering, such as helical channels in quantum spin Hall devices with dilute magnetic impurities or in-plane fields combined with charge disorder, can serve as a suitable and potentially more tunable platform for topological electronics~\cite{Hsu:2025}. 

Quantitative estimates of the relevant energy, disorder, and interaction scales are provided in Appendix~\ref{Appendix:platforms}. For a characteristic bandwidth of $O(10~\mathrm{meV})$, proximity-induced gaps of order $0.1~\mathrm{meV}$ correspond to the dimensionless pairing strengths used in our analysis, while representative mean free paths lead to disorder strengths within the range of our phase diagrams. The interaction parameters are experimentally tunable through electrostatic screening, supporting the accessibility of the predicted phases.

While the numerical analysis above focuses on integer quantum spin Hall edges and the resulting Majorana zero modes, our framework naturally extends to fractional helical liquids~\cite{Levin:2009,Neupert:2011,Santos:2011,Stern:2016,Rachel:2018}. In these systems, proximitized configurations can stabilize parafermionic bound states~\cite{Klinovaja:2014helical}, enabling more advanced schemes for topological quantum computation. The derived RG flow equations here and the additional numerical results presented in Appendix~\ref{Appendix:m3_fractional} indicate a reduced degree of topological protection in the fractional regime, suggesting their enhanced sensitivity to interaction- and disorder-induced fluctuations.
Throughout our numerics, we have considered an edge length of $O(\mu\text{m})$--$O(10~\mu\text{m})$, and the resulting phase diagrams apply for temperatures below $O(0.1~\text{K})$--$O(\text{K})$. For somewhat higher temperatures or shorter edges, similar behavior is expected, with the phase boundaries shifted accordingly.

Beyond transport signatures, the Majorana zero modes stabilized in our setting can be directly probed using local probes. Scanning tunneling spectroscopy with high energy resolution, as demonstrated in the detection of zero-bias bound states in iron-based superconductors~\cite{Machida:2019}, provides an experimentally accessible route to identifying the zero modes predicted here. As revealed by the density-profile analysis, the spatial localization of the zero modes constitutes an alternative scanning-probe feature, with pronounced broadening indicating proximity to a topological phase transition. Overall, the evolution of the local density of states and spatial density profiles under controlled tuning of in-plane fields, interaction strength, or channel asymmetry offers a concrete pathway to experimentally verify the disorder-enabled topological windows, cascades of successive phase transitions, and the revival of Majorana zero modes in future device platforms.

\begin{acknowledgments}

We thank A.~Alexandradinata, B.~A.~Bernevig, Y.-Y.~Chang, A.~Garg, T.~Grover, Y.~Kato, H.-H.~Lu, D.~Parker, S.~Ryu, K.~Shiozaki, S.~Syzranov, H.-C.~Wang, and T.~Yoshida for interesting discussions.  
This work was financially supported by the National Science and Technology Council (NSTC), Taiwan (Grant No.~NSTC-112-2112-M-001-025-MY3, Grant No.~NSTC-114-2112-M-001-057, and Grant No.~NSTC-115-2112-M-001-024), and Academia Sinica (AS), Taiwan (Grant No.~AS-iMATE-114-12).    

\end{acknowledgments}

\section*{Data Availability}

The data that support the findings of this study are available at Zenodo~\cite{data}.

\appendix

\begin{table}[ht]
\caption{Material parameters for semiconductor-based quantum spin Hall insulators.}
\begin{tabular}[c]{  l  c  c }
\hline \hline
Physical parameter & InAs/GaSb~\footnote{From Refs.~\cite{Gueron:1964,Paget:1977,Schliemann:2003,Wu:2006,Braun:2006,Maciejko:2009,Knez:2011,Pribiag:2015,Schrade:2015,Li:2015}.} & HgTe~\footnote{From Refs.~\cite{Konig:2007,Roth:2009,Hou:2009,Maciejko:2009,Strom:2009,Teo:2009,Egger:2010,Lunde:2013}.} \\
\hline
Edge-state Fermi velocity, $v_{F}$ & $4.6 \times 10^4~$m/s & $5.1 \times 10^5~$m/s  \\
Transverse decay length, $a$ & 9~nm & 14~nm\\
Bulk gap, $\Delta=\hbar v_{F} / a$  & $\sim 3$--$35$~meV & $\sim 14$--$55$~meV \\
\hline \hline
\end{tabular}
\label{Tab:semiconductors}
\end{table}

\section{Potential platforms}
\label{Appendix:platforms}

In this section, we compile material parameters of several established quantum spin Hall materials and newly emerging quantum spin Hall systems. These include semiconductor-based materials (see Table~\ref{Tab:semiconductors}), monolayers (see Table~\ref{Tab:monolayer}), and twisted bilayers that host (fractional) quantum spin Hall states (see Table~\ref{Tab:moire}). Conventional semiconductor systems typically have small to moderate bulk gaps and support quantized edge transport at sufficiently low temperatures for small sample size.  
Monolayer systems, such as WTe$_2$, TaIrTe$_4$ and bismuthene on SiC, typically exhibit larger gaps. In particular, TaIrTe$_4$ is notable for hosting a density-tunable dual quantum spin Hall insulating phase~\cite{Tang:2024}, with quantized edge conduction observed up to nearly 100~K. These large-gap and high-temperature systems offer promising conditions for robust helical edge states.

\begin{table*}[hbt]
\caption{Material parameters for monolayer quantum spin Hall insulators.}
\begin{tabular}[c]{ l  c  c  c  c }
\hline \hline
Physical parameter & 1T$^\prime$-WTe$_2$~\footnote{From Refs.~\cite{Tang:2017,Wu:2018,Shi:2019,Jia:2022,Maximenko:2022}.} & Bismuthene on SiC~\footnote{From Refs.~\cite{Reis:2017,Stuhler:2019}.} & TaIrTe$_4$~\footnote{From Refs.~\cite{Xu:2019,Tang:2024}.} \\
\hline
Edge-state Fermi velocity, $v_{F}$ & \begin{tabular}{@{}c@{}} $1.2 \times 10^5~$m/s (Y-edge) \\ $2.7\times 10^5~$m/s (X-edge) \end{tabular} & $5.5 \times 10^5~$m/s &-- \\ 
Transverse decay length, $a$ & 2~nm & 0.4~nm &-- \\
Bulk gap, $\Delta=\hbar v_{F} / a$  & $\sim 55$~meV & $\sim 0.8$~eV & $\sim 10$--$20$~meV \\ 
\hline \hline
\end{tabular}
\label{Tab:monolayer}
\end{table*}

\begin{table}[h]
\caption{Material parameters for moir\'e systems with quantum spin Hall states.
}
\begin{tabular}[c]{ l  c  c }
\hline \hline
Physical parameter & tWSe$_2$ ~\footnote{From Ref.~\cite{Kang:2024b}.} & tMoTe$_2$~\footnote{From Ref.~\cite{Kang:2024a}.} \\
\hline
Bulk gap, $\Delta=\hbar v_{F} / a$  & \begin{tabular}{@{}c@{}} $4$~meV ($\nu=2$) \\ $1.5$~meV ($\nu=4$) \end{tabular} & $0.3$~meV ($\nu=3$) \\
\hline \hline
\end{tabular}
\label{Tab:moire}
\end{table}

\label{Appendix:experimental_estimates}

Here we provide quantitative estimates of the relevant physical parameters and discuss the experimental feasibility of the predicted phases. Motivated by the bulk gap values of the realistic materials discussed above, our numerics employs a characteristic bandwidth $\hbar v_F/a \sim O(10~\mathrm{meV})$. Relative to this scale, a proximity-induced gap of order $0.1~\mathrm{meV}$ corresponds to the representative values $\tilde{\Delta}_n(0),\tilde{\Delta}_c(0)\sim O(10^{-2})$ used in our calculations.
The microscopic disorder strength can be related to the mean free path $\lambda_{\rm mfp}$, whose typical values lie in the range $0.1$--$1~\mu\mathrm{m}$. Taking $\lambda_{\rm mfp}=0.1~\mu\mathrm{m}$ and a Zeeman energy $E_Z\sim O(0.1~\mathrm{meV})$, corresponding to an applied magnetic field of order $1~\mathrm{T}$ for a representative effective $g$ factor, gives an effective spin-flip backscattering amplitude $V_n\sim O(10~\mu\mathrm{eV})$ and a dimensionless disorder strength $\tilde{D}_n\sim O(10^{-6})$. These estimates fall within the parameter ranges considered in our phase diagrams.
The interaction parameters are broadly tunable, and electrostatic screening by a nearby metallic gate can be exploited to tune the interaction parameters over a wide range, approximately $K_n\in[0.1,1]$, as shown in Refs.~\cite{Hsu:2018,HsuPhonons:2024}. 

Together, the estimated pairing, disorder, and interaction strengths lie within the $N_{\rm mzm} \neq 0$ phases identified, for example, in Fig.~\ref{fig:PD1_Nmzm}.
We also remark that our phase diagrams cover broad ranges of the interaction parameter $K_n(0)$ and disorder strength $\tilde{D}_n(0)$, including regimes that are experimentally relevant to several candidate material platforms.

\section{Derivation of random spin-flip terms }
\label{Appendix:rs_terms}

\subsection{Single-particle description }
\label{Appendix:rs_single_particle}

In this section we discuss the origin of the random spin-flip backscattering terms. In practice, it can be induced by several microscopic mechanisms, including magnetic impurities themselves, or a combination of a magnetic field perpendicular to the spin quantization axis and charge disorder. In both cases, the key point is the presence of spin-nonconserving terms with a spatially fluctuating amplitude, which couples right- and left-moving fermions and thereby introduces $2k_F$ backscattering.

\subsubsection{Magnetic impurities} 

Magnetic impurities provide a natural microscopic source of transverse spin-flip backscattering fields for helical channels. In materials such as Mn-doped HgTe quantum wells~\cite{Furdyna:1988,Novik:2005,Dietl:2023a,Dietl:2023b}, the exchange coupling between the dopant moments and the edge states can be substantial~\cite{Liu:2008b,Wang:2014}, producing spatially fluctuating transverse fields that break spin-momentum locking. In the low-energy limit, this coupling can be written in the effective one-dimensional form~\cite{HsuLocalization:2017,HsuTransport:2018},  
\begin{align}
   H_{{\rm rs}}=&  \sum_{\mu , \sigma \sigma^{\prime}} \int dr \; \frac{J^{\mu} }{2} \left[ 
   \Psi^{\dagger}_{\sigma}(r) \sigma^\mu_{\sigma\sigma^{\prime}}
   \Psi_{\sigma^\prime}(r ) \right]   S^\mu (r) ,
\end{align}  
with the exchange coupling $J^{\mu} $ and the impurity spin operator $S^\mu(r)$.
In the helical basis, $(\Psi_{\uparrow}, \Psi_{\downarrow}) = (L_{\uparrow}, R_{\downarrow})$, the electron spin components $\frac 1 2 \Psi^{\dagger}_{\sigma}\sigma^\mu_{\sigma\sigma^{\prime}}
   \Psi_{\sigma^\prime}$ take the form,
\begin{subequations}
\begin{eqnarray}
     & \frac 1 2 \Big[ R_{\downarrow}^\dagger(r) L_{\uparrow}(r) + L_{\uparrow}^\dagger(r) R_{\downarrow}(r) \Big], &\quad  \mu = x,\\
     &\frac i 2 \Big[ R_{\downarrow}^\dagger(r) L_{\uparrow}(r) - L_{\uparrow}^\dagger(r) R_{\downarrow}(r) \Big], & \quad \mu = y,\\
    &\frac 1 2 \Big[ L_{\uparrow}^\dagger(r) L_{\uparrow}(r) - R_{\downarrow}^\dagger(r) R_{\downarrow}(r) \Big], & \quad \mu = z.
\end{eqnarray}
\end{subequations}
Assuming isotropic transverse coupling $J^x=J^y=J$, we can isolate the spin-flip terms by forming linear combinations of the transverse components:
\begin{subequations}
\begin{eqnarray}
    \frac 1 2 \Psi^{\dagger}_{\sigma} (r)(\sigma^x_{\sigma\sigma^{\prime}} + i \sigma^y_{\sigma\sigma^{\prime}}) \Psi_{\sigma^\prime}(r) &=& L^\dagger(r) R(r), \\
    \frac 1 2 \Psi^{\dagger}_{\sigma} (r)(\sigma^x_{\sigma\sigma^{\prime}} - i \sigma^y_{\sigma\sigma^{\prime}}) \Psi_{\sigma^\prime}(r) &=& R^\dagger(r) L(r),
\end{eqnarray}
\end{subequations}
which flip the electron spin and therefore generate backscattering between right- and left-movers. 
One thus obtains 
\begin{equation}
\begin{aligned}
   H_{{\rm rs}}=& \frac 1 2 \int dr \, J \Big[ S^+ (r) R_{\downarrow}^\dagger (r)  L_{\uparrow} (r)  +  S^- (r)  L_{\uparrow}^\dagger (r)  R_{\downarrow} (r) \\
   & +  \cdots \Big],
\end{aligned}
\end{equation}
with $S^\pm (r) \equiv [ S^x (r) \pm i S^y (r)]$ and omitted forward scatterings, which do not contribute to spin-flip backscattering.
After bosonization, one obtains Eq.~\eqref{rs_hamiltonian} in the main text.

\subsubsection{Coexistence of uniform magnetic fields and charge disorder}

Here we discuss coexisting uniform magnetic fields and charge disorder. 
A uniform magnetic field $\mathbf B=(B_x,B_y)$ transverse to the spin quantization axis  couples to the electron spins via
\begin{equation}
    H_B=\frac 1 2\int dr\, \Psi^\dagger(r)(B_x\sigma^x+B_y\sigma^y)\Psi(r).
\end{equation}
While this can flip spins, the spatially constant form carries only the $q=0$ Fourier component and therefore cannot connect the $\pm2k_F$ Fermi points of the right- and left-moving modes.
Therefore, the uniform field cannot generate backscattering by itself.

\begin{figure}[t]
\centering
\includegraphics[width=0.5\linewidth]{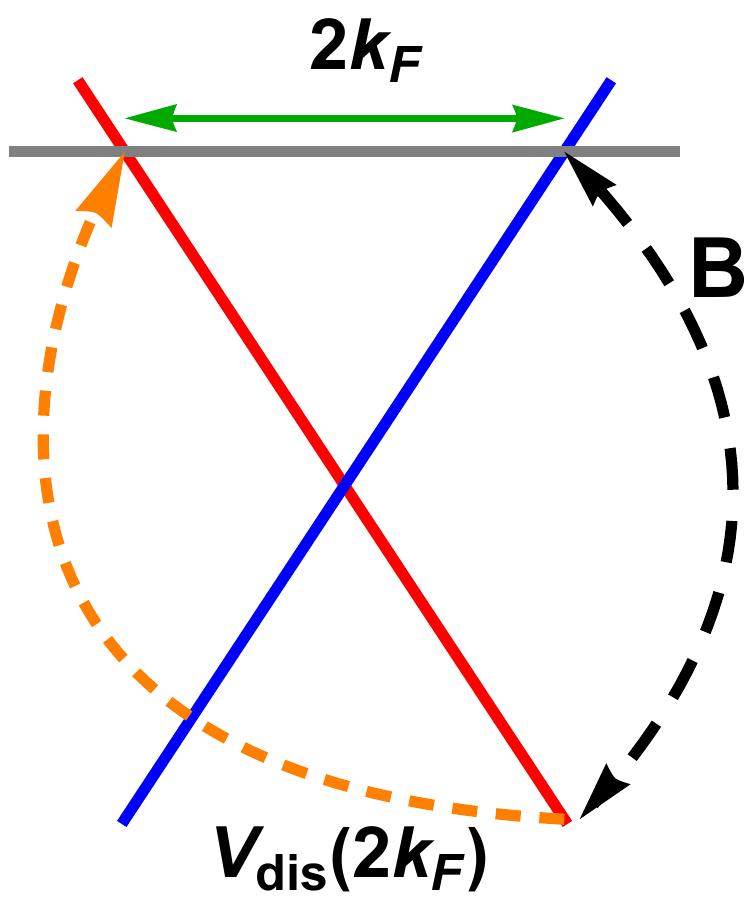}
    \caption{Illustration of spin-flip backscattering process generated by combing a uniform transverse magnetic field $\mathbf B$ and charge disorder potential, $V_{\rm dis}$.    
    }
    \label{fig:charge_imp}
\end{figure}

Nonetheless, the situation changes when we include charge disorder, which produces a random potential $V_{\rm dis}(r)$ and can couple to the electrons in the helical channels.
Specifically, we obtain 
\begin{equation}
    H_{\rm dis} = \int dr \; V_{\rm dis}(r)\rho(r),
\end{equation}
where $\rho(r)$ is the charge density,
\begin{equation}
    \rho(r)=\sum_{n\in\{1,2\}}\Big[ R^\dagger_{n\downarrow}(r)R_{n\downarrow}(r) + L^\dagger_{n\uparrow}(r) L_{n\uparrow}(r) \Big],
\end{equation}
in terms of the right- and left-moving electrons.
We see that $H_{\rm dis}$ only induces spin-conserving forward scattering.

When both of these elements are present, the $2 k_F$ components of $V_{\rm dis}$ supply the momentum difference while $H_B$  provides the spin flip.
We thus get the effective random spin-flip backscattering, which can be recast into
\begin{equation}
    H_{\rm rs} = \sum_{n\in\{1,2\}}\int dr\, \Big[V_{\rm rs,n}(r)R^\dagger_{n\downarrow} L_{n \uparrow} + \text{H.c.} \Big],
\end{equation}
with the effective strength $V_{{\rm rs},n}(r) \propto |\mathbf B| V_{\rm dis} (2 k_{F}) / (\hbar v_{F} k_{F} ) $.
After bosonization, one obtains Eq.~\eqref{rs_hamiltonian} in the main text.
Notably, this realization allows one to externally tune the backscattering strength via external  magnetic fields. We note that Ref.~\cite{Lezmy:2012} studied the scattering processes generated by a single local impurity in a helical liquid.

\subsection{Effective action}
\label{Appendix:rs_effective_action}

To incorporate backscattering into the low-energy theory, we average the partition function over the random fields $V_{{\rm rs},n}(r)$. In the bosonized form, the spin-flip backscattering terms appear as $e^{\pm2mi\phi_n(r)}$. To average over the random fields $V_{{\rm rs},n}(r)$ we follow the standard replica method for disordered Luttinger liquids~\cite{Giamarchi:2004},  where the disorder average is handled by rewriting the inverse functional integral in the average as 
\begin{equation}
    \frac{1}{\int \mathcal D\phi \, e^{-S_V(\phi)}} = \left[\int \mathcal D\phi \, e^{-S_V(\phi)}\right]^{p-1},
\end{equation}
with the limit $p\to 0$ taken at the end. This allows one to express the factor on the right as a product of $p-1$ identical integrals over independent fields $\phi_2,\dots,\phi_p$, after which the average over the disorder can be performed directly. 
Carrying out the average then produces the contribution to the effective action,
\begin{equation}
\begin{aligned}
 &   \frac{\delta S_{\rm rs}}{\hbar}=-\sum _n \frac{\tilde D_n u_n^2}{8 \pi a^3}
    \\
&    \times 
    \int_{u_n |\tau-\tau'|>a} dr d\tau d\tau' 
    \cos[2m\phi_n(r,\tau) -2m\phi_n(r,\tau')],
\end{aligned}
\end{equation}
appearing in Eq.~\eqref{eq:rs_action} in the main text.

In our model, the random backscattering potential $V_{{\rm rs},n}(r)$ is assumed to be a short-range random variable with zero mean. Its strength is characterized by the standard correlator~\cite{Giamarchi:2004}
\begin{equation}
    \overline{\langle V_{{\rm rs}, n}^\dagger(r) V_{{\rm rs}, n'}(r') \rangle} = D_{n}  \delta_{nn'} \delta (r-r').
\end{equation}
One can integrate the correlator over a region of size $a$,
\begin{equation}
\int_{|r-r'|\lesssim a} dr'\;\overline{\langle V_{{\rm rs},n}^\dagger(r) V_{{\rm rs},n}(r')\rangle} \simeq \overline{\langle |V_{{\rm rs},n}(r)|^2\rangle}\; a ,
\end{equation}
and connect the strength $  D_n$ to the random potential,
\begin{equation}
    D_n = a V_n^2,
\end{equation}
with   $V_n^2 \equiv \overline{\langle |V_{{\rm rs},n}(r)|^2\rangle}$ quantifying the backscattering strength.


\section{Derivation of RG flow equations }
\label{Appendix:RG}

In this section, we sketch the derivation of RG flow equations. 
Whereas the following analysis focuses on the single-channel case, it  can be straightforwardly generalized to the double-channel or multiple-channel case~\cite{Hung:2025b}.

To proceed, we give the operator product expansion (OPE) formula used for the derivation.
For a small separation, $z\equiv z_1 - z_2$, around the center-of-mass coordinate, $\displaystyle Z_{\text{c.o.m.}}\equiv ( z_1 + z_2)/2$, one can introduce the currents~\cite{Senechal:1999}
\begin{equation}
\label{current_operators}
\begin{aligned}
J_\phi &\equiv & i\,\partial_z \phi(z,\bar z)\Big|_{(z,\bar z)\to \text{c.o.m.}}, \\
 \bar J_\phi &\equiv& -i\,\partial_{\bar z}\phi(z,\bar z)\Big|_{(z,\bar z)\to \text{c.o.m.}}, \\
J_\theta &\equiv& i\,\partial_z \theta(z,\bar z)\Big|_{(z,\bar z)\to \text{c.o.m.}}, \\
\bar J_\theta &\equiv& -i\,\partial_{\bar z}\theta(z,\bar z)\Big|_{(z,\bar z)\to \text{c.o.m.}}.
\end{aligned}
\end{equation}
With this notation, we get  
\begin{equation}
\label{ope_intermediate_2_currents}
\begin{aligned}
&e^{i(\lambda_\phi \phi(z_1,\bar z_1) + \lambda_\theta \theta(z_1,\bar z_1))} e^{-i(\lambda_\phi \phi(z_2,\bar z_2) + \lambda_\theta \theta(z_2,\bar z_2))} 
\\ &+ e^{-i(\lambda_\phi \phi(z_1,\bar z_1) + \lambda_\theta \theta(z_1,\bar z_1))} e^{i(\lambda_\phi \phi(z_2,\bar z_2) + \lambda_\theta \theta(z_2,\bar z_2))} \\
&\approx \left( 2 - 2\lambda_\phi^2 | z|^2 J_\phi \bar J_\phi
- 2\lambda_\theta^2 | z|^2 J_\theta \bar J_\theta + \dots\right) \\
&\times e^{-\tfrac{1}{2}\langle (\lambda_\phi \phi(z_1,\bar z_1) + \lambda_\theta \theta(z_1,\bar z_1))^2 \rangle_0} 
        e^{-\tfrac{1}{2}\langle (\lambda_\phi \phi(z_2,\bar z_2) + \lambda_\theta \theta(z_2,\bar z_2))^2 \rangle_0} 
        \\ &\times e^{\langle (\lambda_\phi \phi(z_1,\bar z_1) + \lambda_\theta \theta(z_1,\bar z_1))(\lambda_\phi \phi(z_2,\bar z_2) + \lambda_\theta \theta(z_2,\bar z_2)) \rangle_0}.        
\end{aligned}
\end{equation}
Using the standard correlators~\cite{Giamarchi:2004,Senechal:1999}
 \begin{subequations}
\begin{align}
\langle[\phi(z_1,\bar z_1) - \phi(z_2,\bar z_2)]^2\rangle_0 =& K\ln\left( \frac{| z|}a \right), \\
\langle[\theta(z_1,\bar z_1) - \theta(z_2,\bar z_2)]^2\rangle_0 =& \frac 1 K \ln\left( \frac{|  z|}a \right), \\
\lim_{ z_1\to  z_2} \langle {\phi(z_1,\bar z_1)\theta(z_2,\bar z_2)}\rangle_0 =& 0,
\end{align}
\end{subequations}
the OPE formula for the two most singular terms is given by 
\begin{equation}
\label{ope_formula_result}
\begin{aligned}
& e^{i(\lambda_\phi \phi(z_1,\bar z_1) + \lambda_\theta \theta(z_1,\bar z_1))} e^{-i(\lambda_\phi \phi(z_2,\bar z_2) + \lambda_\theta \theta(z_2,\bar z_2))} 
\\&+ e^{-i(\lambda_\phi \phi(z_1,\bar z_1) + \lambda_\theta \theta(z_1,\bar z_1))} e^{i(\lambda_\phi \phi(z_2,\bar z_2) + \lambda_\theta \theta(z_2,\bar z_2))} \\ 
&  \approx  \frac{2}{\left({\frac{|  z|}{a}}\right) ^ {\frac 1 2 \left( \lambda_\phi^2 K + 
\frac{\lambda_\theta^2}{K}\right)}} 
\left( 1 - \frac{(| z| a)^2}{a^2} 
( \lambda_\phi^2 J_\phi \bar J_\phi
+ \lambda_\theta^2 J_\theta \bar J_\theta) + \dots\right).
\end{aligned} 
\end{equation}
 
For a general perturbation form,
\label{Appendix:RG_flow}
\begin{equation}
    \delta H = \frac {g}{\pi a}\int dx \cos(\lambda_\phi \phi + \lambda_\theta \theta),
\end{equation}
rescaling the cutoff $a\rightarrow a(1+dl)$ gives the flow of the dimensionless coupling, 
\begin{equation}
        \frac{d \tilde g}{dl} = \left[\ 2 - \frac 1 4 \left( \lambda_\phi ^2 K + \frac{\lambda_\theta^2}{K} \right) \right] \tilde g.
\end{equation}
For the second-order term, one obtains   
\begin{equation}
\begin{aligned}
   & \frac{1}{2}\left\langle \left(\frac{\delta S}{\hbar}\right)^2 \right\rangle_0 = 
   \frac 1 8 \left( \frac{\tilde g}{\pi a^2} \right)^2 
   \\ & \times \int d^2 \mathbf x _1 d^2 \mathbf x _2 \sum_{\epsilon_1,\epsilon_2 } \left\langle e^{i\epsilon_1(\lambda_\phi \phi(\mathbf x_1) + \lambda_\theta \theta(\mathbf x_1))} 
   \right.
   \\ &\qquad\qquad\qquad\qquad\quad \left.\times
   e^{i\epsilon_2(\lambda_\phi \phi(\mathbf x_2) + \lambda_\theta \theta(\mathbf x_2))} \right\rangle_0,
\end{aligned}
\end{equation} 
where $\epsilon_1,\epsilon_2 \in \{+,-\}$. The terms with opposite signs in the exponents ($\epsilon_1=-\epsilon_2$) give the most singular contributions. For these terms, we apply the OPE relation in Eq.~\eqref{ope_formula_result}. Changing the coordinates to ($\mathbf X, \mathbf x$), where $\mathbf x\equiv \mathbf x_1-\mathbf x_2$ and $\mathbf X\equiv (\mathbf x_1+\mathbf x_2)/2$, and averaging the fast coordinate over the shell $a\le |\mathbf  x|\le a(1+dl)$ renormalizes the quadratic action through the $J_\phi \bar J_\phi$ and $J_\theta \bar J_\theta$ terms, resulting in
\begin{equation}
    \frac{dK}{dl}=\left( \frac{\lambda_\theta^2}4 - \frac{\lambda_\phi^2}{4}K^2 \right) \tilde g^2.
\end{equation}
The above results can be used to derive the contributions of the pairing terms to the final RG flow equations.

We also discuss the RG contributions from the   $\tilde D_n$ terms for a general coefficient $\lambda_\phi$. As in Refs.~\cite{Giamarchi:1988,Giamarchi:2004}, the flow is obtained by studying the correction to the correlator
\begin{equation}
\begin{aligned}
    R_n(\mathbf r_1-\mathbf r_2)\equiv& \big\langle e^{i[\phi_n(\mathbf r_1)-\phi_n(\mathbf r_2)]}\big\rangle_{S_{{\rm el},n}+S_{{\rm rs},n}}\\
    &=\frac{1}{\mathcal Z_n} \int \mathcal D\phi_n e^{-S_{{\rm el},n}/\hbar}e^{-S_{{\rm rs},n}/\hbar} e^{i[\phi_n(\mathbf r_1)-\phi_n(\mathbf r_2)]}, 
\end{aligned}
\end{equation}
with $\mathbf r_{1,2}\equiv (r_{1,2},y_{1,2})$, $y=u_n\tau$ and the partition function $\mathcal Z_n\equiv \int \mathcal D\phi_n  e^{-(S_{{\rm el},n}+S_{{\rm rs},n})/\hbar}$. Under the rescaling, the correlator preserves the helical liquid form with renormalized parameters.
In the above, the pairing terms are omitted in $\mathcal Z_n$ as the crossed-term contributions to the RG flow vanish; see Appendix~\ref{Appendix:RG_cross}. 

\begin{widetext}
The action consists of two parts,
\begin{subequations}
\begin{eqnarray}
    \frac{S_{{\rm el},n}}\hbar &=& \int dr\,d\tau\frac{1}{2\pi K_n}\left[\frac{1}{u_n}  (\partial_\tau \phi_n(r,\tau))^2 +u_n (\partial_r \phi_n(r,\tau))^2 \right], \nonumber\\
    \\
    \frac{S_{{\rm rs},n}}\hbar &=& -\frac{D_n}{(2\pi \hbar a)^2 u_n^2}  
    \int_{|y-y'|>a} dr dy dy' 
    \cos[\lambda_\phi \phi_n(r,\tau) - \lambda_\phi \phi_n(r,\tau')]. 
\end{eqnarray}
\end{subequations}
We then obtain
\begin{equation}
\begin{aligned}
    R_n(\mathbf r_1-\mathbf r_2)&\approx \big\langle e^{i[\phi_n(\mathbf r_1)-\phi_n(\mathbf r_2)]}\big \rangle_0\\
    &+\frac{D_n}{(2\pi \hbar a)^2 u_n^2}\int_{|y-y'|>a} dr\,dy\,dy'\, \big\{\big\langle e^{i[\phi_n(\mathbf r_1) - \phi_n(\mathbf r_2)]} \cos(\lambda_\phi \phi_n(r,\tau) - \lambda_\phi \phi_n(r,\tau')) \big\rangle_0\\
    &\qquad\qquad\qquad\qquad\qquad\qquad\quad - \big\langle e^{i[\phi_n(\mathbf r_1) - \phi_n(\mathbf r_2)]} \big\rangle_0 \big\langle \cos(\lambda_\phi \phi_n(r,\tau) - \lambda_\phi \phi_n(r,\tau')) \big\rangle_0
    \big\}.
\end{aligned}
\end{equation}
\end{widetext}
The zeroth-order correlator is
\begin{equation}
    \big\langle e^{i[\phi_n(\mathbf r_1)-\phi_n(\mathbf r_2)]} \big\rangle_0 = e^{ - \frac 1 2 K_n F(\mathbf r_1-\mathbf r_2)},
\end{equation}
where
\begin{equation}
    \big\langle [\phi_n(\mathbf r_1)-\phi_n(\mathbf r_2)]^2 \big\rangle_0 = K_n F(\mathbf r_1 - \mathbf r_2).
\end{equation}
Combining it with the first order terms, the correlator preserves the same form but with renormalized coefficients,
\begin{equation}
\begin{aligned}
    R_n(\mathbf r_1 - \mathbf r_2) = & \exp\left[ -\frac{K_{{\rm eff},n}} 4 \ln\left( \frac{(\mathbf r_1 - \mathbf r_2)^2}{a^2} \right) \right. \\  & \qquad \qquad -\left.\frac{t_{\perp,{\rm eff},n}}{2} \cos(2\theta_{\mathbf r_1-\mathbf r_2}) \right].
\end{aligned}
\end{equation} 
Here, $\theta_{\mathbf r_1 - \mathbf r_2}$ is the angle between the vector $\mathbf r_1 - \mathbf r_2$ and the $r$ axis, $t_{\perp,n}$ parametrizes the anisotropy between the spatial and temporal coordinates for channel $n$, and 
\begin{subequations}
\begin{eqnarray}
    K_{{\rm eff},n} &\equiv& K_n - \frac{\lambda_\phi^2 K_n^2}{8}\tilde D_n \int_a^\infty \frac{dz}{a}\left(\frac z a\right)^{2-\frac{\lambda_\phi^2 K_n}{2}}, \nonumber \\ 
    t_{\perp,{\rm eff},n} &\equiv& t_{\perp,n}+\frac{\lambda_\phi^2 K_n^2}{16}\tilde D_n \int_a^\infty \frac{dz}{a}\left(\frac z a\right)^{2-\frac{\lambda_\phi^2 K_n}{2}}. \nonumber 
\end{eqnarray}
\end{subequations}
Thus we obtain the flow equations
\begin{subequations}
\begin{eqnarray}
    \frac{d \tilde D_n}{dl} &=& \left[ 3 -\frac{\lambda_\phi^2}{2}K_n \right] \tilde D_n,\\
    \frac{d K_n}{dl} &=& -\frac{\lambda_\phi^2 K_n^2}{8}\tilde D_n,\\
    \frac{d t_{\perp,n}}{dl} &=& \frac{\lambda_\phi^2 K_n^2}{16} \tilde D_n.
\end{eqnarray}
\end{subequations}
Relating $t_{\perp,n}$ and the velocity $u_n$ as
\begin{equation}
    \frac {d u_n}{dl} = -\frac{2 u_n}{K_n} \frac{d t_{\perp,n}}{dl}
\end{equation}
further gives
\begin{equation}
    \frac{du_n}{dl} = -\frac {\lambda_\phi^2 u_n K_n}{8} \tilde D_n.
\end{equation}
We utilize the derived relations here to include contributions from $D_{n}$ terms in the final RG flow equations presented in the main text. 

\subsection{RG procedure for general $m$}
\label{Appendix:RG_fractional}
The fractional commutation relation in Eq.~\eqref{fields_commutation_relation} can be brought to the standard form by rescaling the fields, 
\begin{align}
    \tilde \phi_n = \sqrt m \phi_n,\quad
    \tilde \theta_n = \sqrt m \theta_n,
\end{align}
so that $[\tilde \phi_n(r), \tilde \theta_{n'}(r')]=\frac{i\pi}{2} \delta_{nn'}\text{sign}(r'-r)$. In this rescaled basis, the pairing terms acquire a factor of $\sqrt m$, for example $\cos(2m\theta_n)$ transforms into $\cos(2\sqrt m \tilde \theta_n)$, with analogous substitutions for the remaining terms. The quadratic action retains its form if the velocity is rescaled as
\begin{equation}
    u_n\rightarrow\frac{u_n}m.
\end{equation}
With these substitutions, the analysis in Appendix~\ref{Appendix:RG} applies directly to the tilded fields, and one can derive the RG flow equations accordingly.

\subsection{Contributions from the crossed terms}
\label{Appendix:RG_cross}

We now consider the second-order expansion of the partition function,
\begin{equation}
    \mathcal Z = \mathcal Z_0\left[ 1 - \left\langle \frac{\delta S}{\hbar} \right\rangle_0 + \frac{1}{2} \left\langle \left( \frac{\delta S}{\hbar} \right)^2 \right\rangle_0 + \dots \right],
\end{equation}
where $\delta S = S_{s,1} + S_{s,2} + S_{\times} + \dots$, corresponding to the non-quadratic terms in the effective action.
The contributions from the non-crossed terms, such as $\langle S_{\mathrm{s},1}^2/\hbar^2\rangle_0$, have been discussed above.
Here we explicitly examine the crossed terms between the local and nonlocal pairings,
\begin{equation}
    \begin{aligned}
         &\frac{S_{\mathrm{s},1}}{\hbar} = \frac{\tilde \Delta_1}{\pi a^2}\int d^2\mathbf x\, \cos(2\sqrt m\tilde \theta_1),\\
         &\frac{S_\times}{\hbar} = \frac{\tilde \Delta_c}{\pi a^2} \int d^2\mathbf x \left\{ \cos\left[\sqrt{m}(-\tilde \phi_1+\tilde\theta_1 + \tilde\phi_2 + \tilde \theta_2)\right]\right. \\
         &\qquad\qquad\qquad\qquad\left.+ \cos\left[\sqrt{m}(\tilde \phi_1+\tilde\theta_1 - \tilde\phi_2 + \tilde \theta_2)\right] \right\}.
    \end{aligned}
\end{equation}
We first evaluate the OPE for the contraction of $S_{\mathrm{s},1}$ with the first term of $S_\times$. In contrast to Eq.~\eqref{ope_formula_result}, the field mismatch in the crossed terms yields a global prefactor,
\begin{equation}
\label{cross_terms_ope_formula_result}
\begin{aligned}
& e^{i 2\sqrt m \tilde \theta_1(z_1,\bar z_1)} e^{-i\sqrt m(- \tilde\phi_1(z_2,\bar z_2) + \tilde\theta_1(z_2,\bar z_2) + \tilde\phi_2(z_2,\bar z_2) + \tilde\theta_2(z_2,\bar z_2))} 
\\&+ e^{-i 2\sqrt m \tilde \theta_1(z_1,\bar z_1)} e^{i\sqrt m(-\tilde\phi_1(z_2,\bar z_2) + \tilde\theta_1(z_2,\bar z_2) + \tilde\phi_2(z_2,\bar z_2) + \tilde\theta_2(z_2,\bar z_2))}   \\ 
&  \approx  \frac{2}{ \left(\frac L a \right)^{\frac m 4\left( K_1 + \frac 1 {K_1} + K_2 + \frac 1 {K_2} \right)} \left(\frac {|z|} a \right)^{\frac m {K_1}}} \\ & \quad \times  \left( 1 - \frac{| z|^2}{a^2} a^2( \frac{9m}{4} J_\phi \bar J_\phi
+ \frac{m}{4} J_\theta \bar J_\theta) + \dots\right),
\end{aligned} 
\end{equation}
 dependent on the system size $L$. 
To find the contribution to the RG flow of $K_1$, we integrate the relative coordinate over the thin shell $a\le |z|\le a(1+dl)$, leading to 
\begin{equation}
    \frac{dK_1}{dl}\propto \left(\frac L a\right) ^{- \frac {m}{4}(K_1 + \frac 1{K_1} + K_2 + \frac 1 {K_2})} \tilde \Delta_1\tilde \Delta_c.
\end{equation}
Since the scaling exponent is strictly positive, this RG contribution is strongly suppressed for sufficiently large $L/a$. Similarly, the other crossed term combinations are also suppressed and the RG flow of $K_n$ is dominated by the non-crossed-term contributions presented in Eq.~\eqref{eqn:rg_flow_K}. Similarly, crossed-term corrections to the couplings, such as  $d\tilde \Delta_c/dl \propto \tilde\Delta_1 \tilde \Delta_c$) are also strongly suppressed.

\section{Detailed analysis of the topological properties }
\label{Appendix:single-particle-analysis}

\subsection{ Symmetry class of the effective Hamiltonian }
\label{Sec:sp-Hamiltonian}
 
In the main text we define the effective Hamiltonian in the single-particle description, where the Hamiltonian density is expressed as 
\begin{equation}
\label{Hspinflip_nonzero_By}
    \mathcal{H}_{\rm sp}=\mathcal{H}_0 + \mathcal{H}_{\rm s} + \mathcal{H}_{ \times} +   \mathcal{H}_{\rm sf}.
\end{equation}
Here, $\mathcal{H}_0$ corresponds to the kinetic energy, 
$\mathcal{H}_\text{\rm s}$ describes local   pairing, $\mathcal{H}_{\times}$ describes the nonlocal pairing, and $\mathcal{H}_{\rm sf}$ corresponds to the spin-flip backscattering terms. Specifically, we have 
\begin{subequations}
\begin{eqnarray}
    \mathcal{H}_0 &=& \hbar v_F k \eta^0 \tau^0 \sigma^z, \\
    \mathcal{H}_\text{\rm s} &=& - \Delta_+ \eta^y \tau^0 \sigma^y - \Delta_- \eta^y \tau^z \sigma^y, \\
    \mathcal{H}_\times &=& - \bar\Delta_c \eta^y \tau^x \sigma^y, \\
    \mathcal{H}_{\rm sf} &=& B_+ \eta^z \tau^0  \sigma^x + B_-\eta^z  \tau^z  \sigma^x , 
\end{eqnarray}    
\end{subequations}
where $\eta^{\mu}$, $\tau^{\mu}$, and $\sigma^{\mu}$ with $\mu \in \{ 0, x, y, z \}$ are identity and Pauli matrices acting on particle-hole, channel, and spin subspaces, respectively ($\mu = {0}$ corresponding to the identity matrix).
Here we denote the spin flip terms as $B_{\pm}$, taken to be real, which represents a general source that can arise from either uniform fields or the root-mean-square amplitude of random fields. 

It can be shown that  $\mathcal{H}_{\rm sp}$ cannot be block-diagonalized and is thus irreducible
to motivate the tenfold classification analysis~\cite{Schnyder:2008,Ryu:2010} of its antiunitary symmetries.  
Despite the presence of the spin-flip terms, the system retains an effective time-reversal symmetry. Specifically, it satisfies
\begin{equation}
\mathcal T \mathcal H_{\mathrm{sp}}(k) \mathcal T^{-1} = \mathcal H_{\mathrm{sp}}(-k),
\end{equation}
with $\mathcal T = U_T \mathcal K$, the complex conjugation $\mathcal K$ and $U_T = \eta^{y}\tau^{0}\sigma^{y}$ or $U_T = \eta^{z}\tau^{0}\sigma^{x}$. 
We thus have $\mathcal T^{2}=+1$.  

The Hamiltonian density also possesses particle-hole symmetry,
\begin{equation}
\mathcal C \mathcal H_{\mathrm{sp}}(k) \mathcal C^{-1} = -\mathcal H_{\mathrm{sp}}(-k),
\end{equation}
where $\mathcal C = U_C \mathcal K$ is an antiunitary particle-hole operator. One may take $U_C = \eta^{0}\tau^{0}\sigma^{z}$ or $U_C = \eta^{x}\tau^{0}\sigma^{0}$, both yielding $\mathcal C^{2}=+1$.

In addition, the system exhibits chiral symmetry defined by the unitary operator $\mathcal S = \mathcal T \mathcal C$. We thus identify the Hamiltonian as belonging to the BDI symmetry class. In one dimension, it supports a $\mathbb{Z}$  invariant that counts the number of Majorana zero modes localized at the system boundaries~\cite{Schnyder:2008,Ryu:2010}.

\subsection{ Majorana zero modes and topological criterion }
\label{Sec:sp-top}
 
Before searching for Majorana zero modes, we look into the ``bulk'' energy spectrum by assuming translational invariance along the channels and obtain 
\begin{widetext}
    \begin{equation}
\label{bulk_energy_spectrum}
    E_{\lambda,\varepsilon}^{\pm} (k) = \pm \sqrt{ \hbar^2 v_F^2 k^2
    +
    \left[(\Delta_+ + \varepsilon B_+) +\lambda\sqrt{\left(\Delta_- +\varepsilon B_- \right)^2+\bar\Delta_c^2} \right]^2  },
\end{equation}
\end{widetext}
where $\lambda,\varepsilon\in\{+,-\}$ label the four distinct energy bands. 
The bulk gap closes at momentum $k=0$ under the following condition,
\begin{equation}
\label{gap_closing_condition}
    \left( \Delta_+ + \varepsilon B_+\right) + \lambda\sqrt{\left(\Delta_- +\varepsilon B_- \right)^2+\bar \Delta_c^2}=0, 
\end{equation}
at which a topological phase transition may occur and the number of zero modes can change. 
 
To search for zero energy modes, we consider the real space setup $H_{\mathrm{sp}}(k) \to H_{\rm sp} (r)$  near a system corner at $ r=0$, which separates the region with spatially dependent nonlocal pairing amplitude $\bar{\Delta}_c\ne0$ ($ r >0$) and the region $\bar{\Delta}_c=0$ ($ r <0$)~\cite{Klinovaja:2014helical,Hsu:2015}. 
We thus have uniform local pairing amplitudes $\Delta_n$ and a spatially dependent nonlocal pairing, $\bar{\Delta}_c (r) = \Delta_{c} \Theta (r)$.
The Majorana  zero-energy solutions satisfy  the Bogoliubov-de Gennes equation, 
\begin{equation}
    \mathcal H_{\rm sp} (r) \Phi_{\rm mzm} (r) =0,
    \label{Eq:BdG}
\end{equation}
subject to the self-conjugation condition and continuity at $r=0$.
We then describe a zero-energy bound state by an eight-component Nambu spinor. Since the system parameters differ on the two sides of the boundary, we write  wavefunctions separately for $r>0$ and $r<0$,
\begin{equation}
    \Phi_{\rm mzm}(r)=\begin{cases}
        \Phi_{\rm mzm}^<(r), \quad \text{ for $r<0$},\\[3pt]
        \Phi_{\rm mzm}^>(r), \quad \text{ for $r>0$},
    \end{cases}
\end{equation}
which take the form, 
\begin{subequations}
\label{Eq:mzm-ansatz}
\begin{eqnarray}
    \Phi_{\text{mzm}}^{<}( r) &=& 
    \begin{pmatrix}
        A^{<} \\
        B^{<} \\
        C^{<} \\
        D^{<} \\
        {A^{<}}^* \\
        {B^{<}}^* \\
        {C^{<}}^* \\
        {D^{<}}^* 
    \end{pmatrix} 
        e^{ \kappa^<r}, \quad \text{ for $r < 0$} , \label{eq:Phi_less}\\[4pt] 
    \Phi_{\text{mzm}}^{>}( r) &=& 
    \begin{pmatrix}
        A^{>} \\
        B^{>} \\
        C^{>} \\
        D^{>} \\
        {A^{>}}^* \\
        {B^{>}}^* \\
        {C^{>}}^* \\
        {D^{>}}^* 
    \end{pmatrix} 
        e^{-\kappa^>r}, \quad \text{ for $r > 0$}. \label{eq:Phi_greater} 
\end{eqnarray}
\end{subequations}
In the above, the coefficients are collected into the spinors  
in the right/left-moving basis 
$ (R_{1\downarrow},L_{1\uparrow},R_{2\downarrow},L_{2\uparrow},R_{1\downarrow}^\dagger,L_{1\uparrow}^\dagger,R_{2\downarrow}^\dagger,L_{2\uparrow}^\dagger)^T$ with the transpose operator $T$. 
For localized modes, the wavefunction takes an exponentially decaying form on both sides of $r=0$, with real decay constants  $\kappa^>,\kappa^< > 0$.

With the above ansatz, we solve Eq.~\eqref{Eq:BdG} and obtain four independent decaying basis states on either side, which are used to match the boundary condition at $r=0$.  
In each region, the general Majorana wavefunction can be expressed as a linear combination of the corresponding decaying basis states, 
\begin{subequations}
\label{eq:phi_expansions}
\begin{eqnarray}
    \Phi^<_{\rm mzm} (r) &=& \sum _{j=1}^{4} c_j \Phi_j^< (r), \quad \text{ for $r<0$}, \\
    \Phi^>_{\rm mzm} (r) &=& \sum _{j=1}^{4} d_j \Phi_j^> (r), \quad \text{ for $r>0$}. 
\end{eqnarray}
\end{subequations}
The decaying basis functions in the two regions are written as 
\begin{subequations}
\begin{eqnarray}
    \Phi_j^{<}(r) & \equiv & \Psi_{j}^{<}e^{ \kappa^{<}_j r}, \quad \text{ for $r<0$}, \label{eq:Phi_less_def}\\ [4pt]
    \Phi_j^{>}(r) & \equiv & \Psi_j^{>}e^{-\kappa^{>}_j r}, \quad \text{ for $r>0$}.\label{eq:Phi_greater_def}
\end{eqnarray}
\end{subequations}
With these preliminaries, we now list the decaying eigenstates used in the boundary matching.

For $r<0$, each channel yields two decaying modes labeled by $\varepsilon\in \{+,-\}$, with decay constants
\begin{equation}
\label{eq:kappa_less}
  \kappa_{n,\varepsilon}^{<} = 
  \frac{1}{\hbar v_F}\,\big|\Delta_n + \varepsilon B_{x,n}\big|,
\end{equation}
and the upper components of the corresponding eigenvectors, 
\begin{widetext} 
\begin{equation}
\begin{aligned}
\Psi_{n,+}^{<}  = 
\begin{pmatrix}
  i \, \delta_{n,1} \\[2pt]
  -\,\mathrm{sgn}(\Delta_n + B_{x,n}) \, \delta_{n,1} \\
  i \, \delta_{n,2} \\[2pt] 
  -\,\mathrm{sgn}(\Delta_n + B_{x,n}) \, \delta_{n,2}
\end{pmatrix}, \; \; 
\end{aligned}
\begin{aligned}
\Psi_{n,-}^{<}  = 
\begin{pmatrix}
  \delta_{n,1} \\[2pt]
  -\,i\,\mathrm{sgn}(\Delta_n - B_{x,n}) \, \delta_{n,1} \\
  \delta_{n,2} \\[2pt] 
  -\,i\,\mathrm{sgn}(\Delta_n - B_{x,n}) \, \delta_{n,2}
\end{pmatrix}, 
\end{aligned}
\end{equation}
\end{widetext} 
up to normalization prefactors.
In the above, the four solutions with $(n,\epsilon)$ for $n \in \{ 1,2 \}$ and $\epsilon \in \{+, - \}$  correspond to the four $\Psi_{j}^{<}$'s for $j \in \{1, \cdots, 4 \}$ in Eq.~\eqref{eq:Phi_less_def}.
The lower Nambu components follow from particle-hole self-conjugation, as enforced in Eq.~\eqref{Eq:mzm-ansatz}, and are therefore not shown.

\begin{figure*}[ht]
\centering
\includegraphics[width=1.0\linewidth]{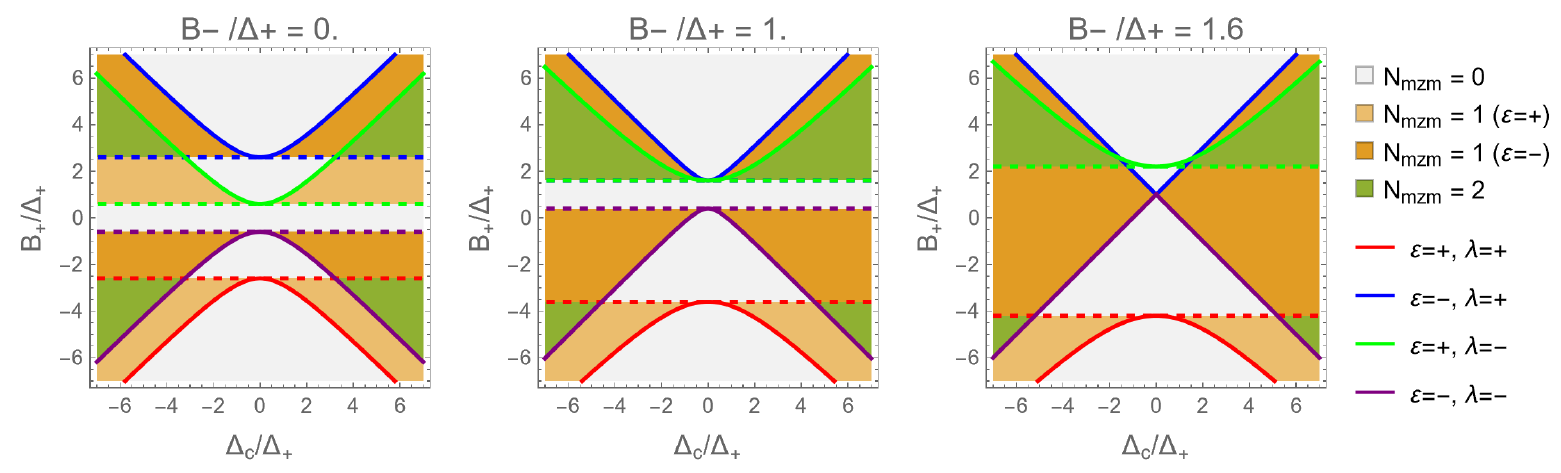}
    \caption{Topological phase diagram constructed based on $N_{\text{mzm}}$ as a function of  $\Delta_c/\Delta_+$ and   $B_+/\Delta_+$. Colors indicate $N_{\text{mzm}}$ as computed from  Eq.~\eqref{eq:NmzmAppendix} and verified numerically; regions with $N_{\text{mzm}}=1$ are further resolved according to which of the $\varepsilon$ conditions in Eq.~\eqref{eq:Nmzm_equality_appendix} is fulfilled.
    The solid (curved) and dashed (horizontal) lines coincide with the analytical phase boundaries given in Eq.~\eqref{gap_closing_condition} of the coupled ($\bar \Delta_c\ne 0$) and decoupled ($\bar \Delta_c=0$) regions, respectively.
    }
    \label{fig:PD_single_particle}
\end{figure*}

For $r>0$, we obtain four decaying modes
labeled by  $\lambda,\varepsilon\in\{+,-\}$. Their decay
constants are
\begin{equation}
\label{eq:kappa_more}
  \kappa_{\lambda,\varepsilon}^{>} = \frac{1}{\hbar v_F}
  \left|(\Delta_+ + \varepsilon B_+) +
  \lambda \sqrt{(\Delta_- + \varepsilon B_-)^2 + \Delta_c^2}\right|,
\end{equation}
and the upper components of the corresponding eigenvectors read
\begin{widetext} 
\begin{eqnarray}
\Psi_{\lambda,+}^{>} &= 
\begin{pmatrix}
  i \\
  s_{\lambda,+} \\
  -\dfrac{i}{\Delta_c}\!\left[(\Delta_- + B_-)  - \lambda\sqrt{(\Delta_- + B_-)^2 + \Delta_c^2}\right] \\
  -s_{\lambda,+}\,\dfrac{1}{\Delta_c}\!\left[(\Delta_- + B_-) - \lambda\sqrt{(\Delta_- + B_-)^2 + \Delta_c^2}\right]
\end{pmatrix},  
\,
\Psi_{\lambda,-}^{>} &= 
\begin{pmatrix}
  1 \\
  i\,s_{\lambda,-} \\
  -\dfrac{1}{\Delta_c}\!\left[(\Delta_- - B_-) -\lambda\sqrt{(\Delta_- - B_-)^2 + \Delta_c^2}\right] \\
  -s_{\lambda,-}\,\dfrac{i}{\Delta_c}\!\left[(\Delta_- - B_-) - \lambda\sqrt{(\Delta_- - B_-)^2 + \Delta_c^2}\right]
\end{pmatrix},   \nonumber \\
\end{eqnarray}     
\end{widetext}
with a $\lambda$- and $\varepsilon$-dependent sign,
\begin{equation}
  s_{\lambda,\varepsilon} \equiv
  \operatorname{sgn}\!\left[(\Delta_+ + \varepsilon B_+) +
  \lambda \sqrt{(\Delta_- + \varepsilon B_-)^2 + \Delta_c^2}\right].
\end{equation}
In the above, the four solutions with $(\lambda,\epsilon)$ for $\lambda, \epsilon \in \{+, - \}$  correspond to the four $\Psi_{j}^{>}$'s for $j \in \{1, \cdots, 4 \}$ in Eq.~\eqref{eq:Phi_greater_def}.

With the  limits ($\Phi_j^{<}$ and $\Phi_j^{>}$) of the decaying basis states toward $r=0$ from the two sides, we can match the boundary condition,  
$     \Phi^{<}_{\rm mzm}(0)=\Phi^{>}_{\rm mzm}(0).$  
With Eq.~\eqref{eq:phi_expansions}, we have 
\begin{equation}
    \sum_{j=1}^4 c_j\,\Psi_j^{<}
    = 
    \sum_{j=1}^4 d_j\,\Psi_j^{>},
\end{equation}
from which we find the number $N_{\rm mzm}$ of Majorana zero modes. 
A change in $N_{\rm mzm}$ in the parameter space requires a bulk gap closing, which is also a condition when a decaying mode disappears. 

With the above formulation, we numerically evaluate $N_{\mathrm{mzm}}$ across parameter space to construct the topological phase diagram within the single-particle description. 
A series of examples is given in Fig.~\ref{fig:PD_single_particle}. 
The phase boundaries coincide precisely with the bulk gap-closing conditions in Eq.~\eqref{gap_closing_condition}, with straight lines arising from the decoupled region ($\bar \Delta_c=0$) and curved lines from the coupled region ($\bar \Delta_c\neq 0$). Thus, the full phase diagram is determined by the superposition of the phase boundaries from both regions. The topological invariant $N_{\text{mzm}}$ can change only when one of these boundaries is crossed.
 
In addition to the numerical evaluation, we derive an equivalent but more compact analytical formula from Eq.~\eqref{gap_closing_condition}.
 This gives 
 \begin{equation}
\label{eq:NmzmAppendix}
\begin{aligned}
    N_{\text{mzm}} = \sum_{\varepsilon \in \{+,-\}} \Theta \left( - |\Delta_- + \varepsilon B_-| + |\Delta_+ + \varepsilon B_+| \right) \\ \times \Theta (  \sqrt{(\Delta_- + \varepsilon B_- )^2 + \Delta_c^2} - |\Delta_+ + \varepsilon B_+| ) , 
\end{aligned}
\end{equation}
which will be used in the main text. 
For a given $\varepsilon$, one zero mode emerges when the inequality 
\begin{equation} 
\label{eq:Nmzm_equality_appendix}
    |\Delta_- + \varepsilon B_-| < |\Delta_+ + \varepsilon B_+| < \sqrt{(\Delta_- + \varepsilon B_-)^2 + \Delta_c^2}
\end{equation}
is satisfied. Thus, for each $\varepsilon$ value, Eq.~\eqref{eq:NmzmAppendix} defines a pair of bounds, 
\begin{subequations}
\begin{eqnarray}
    L_\varepsilon & \equiv & |\Delta_-+\varepsilon B_-|,\\
    U_\varepsilon & \equiv & \sqrt{(\Delta_- + \varepsilon B_-)^2 + \Delta_c^2}.
\end{eqnarray}
\end{subequations}
For a given $\varepsilon$, the quantity $|\Delta_+ + \varepsilon B_+|$ must fall between these two bounds in order to generate a zero mode. In the backscattering-free limit ($B_\pm = 0$), the two sectors share identical bounds, $L_+ = L_- = |\Delta_-|$ and $U_+=U_-=\sqrt{\Delta_-^2 + \Delta_c^2}$, so both conditions collapse to the same inequality. This forces the two Majorana zero mode contributions to appear or disappear together.

The presence of the backscattering terms breaks this locking, which we extensively explore in this work.
An instructive special case is symmetric backscattering, with $B_{+}\neq 0$ and $B_-=0$. Here the upper and lower bounds are identical for both $\varepsilon$ sectors, with different middle terms, $|\Delta_+ + \varepsilon B_+|$. Thus, this is sufficient to lift the degeneracy between the two Majorana zero mode conditions, allowing a single zero mode ($N_{\rm mzm}=1$) to appear--a regime that cannot be achieved in the clean system even with asymmetric pairing, $\Delta_-\ne0$.

\begin{table}[h!]
\centering
\caption{Parameter sets adopted for the numerics of our RG analysis. }
\begin{tabular}{l c c c c c c c}
\hline
\hline 
\vspace{-8pt}       &  & &   &  &   &  &   \\
 \textbf{Figure}        & $\tilde \Delta_+(0) $ & $\tilde \Delta_-(0)$ & $\tilde\Delta_c(0)$ & $\tilde D_+(0)$ & $\tilde D_-(0)$ & $ K_{+} (0) $ & $K_-(0) $ \\
\hline 
Fig.~\ref{fig:pointPflow}(a)                       & 0.03 & 0.01  & 0.01  & $10^{-6}$ & 0         & 0.57      & 0    \\ 
Fig.~\ref{fig:pointPflow}(b--f)                    & 0.03 & 0.01  & 0.01  & --        & 0         & --        & 0    \\ 
\hline 

Fig.~\ref{fig:PD_transport1}(a)                    & 0.03 & 0     & 0.01  & --        & 0         & --        & 0    \\ 
Fig.~\ref{fig:PD_transport1}(b)                    & 0.03 & 0.01  & 0.01  & --        & 0         & --        & 0    \\ 
Fig.~\ref{fig:PD_transport1}(c)                    & 0.03 & 0     & 0.01  & --        & 0         & --        & 0.05 \\ 
\hline
Fig.~\ref{fig:localization_length_temperature}     & 0.03 & 0.01  & 0.01  & --        & 0         & --        & 0    \\ 
\hline
Fig.~\ref{fig:PD1_Nmzm}(a)                         & 0.03 & 0     & 0.01  & --        & 0         & --        & 0    \\ 
Fig.~\ref{fig:PD1_Nmzm}(b)                         & 0.03 & 0.01  & 0.01  & --        & 0         & --        & 0    \\ 
Fig.~\ref{fig:PD1_Nmzm}(c)                         & 0.03 & 0     & 0.01  & --        & 0         & --        & 0.05 \\ 
\hline
Fig.~\ref{fig:flow_I_to_V}(a)                      & 0.03 & 0     & 0.01  & $10^{-6}$ & 0         & 0.43      & 0.05 \\
Fig.~\ref{fig:flow_I_to_V}(b)                      & 0.03 & 0     & 0.01  & $10^{-6}$ & 0         & 0.5       & 0.05 \\
Fig.~\ref{fig:flow_I_to_V}(c)                      & 0.03 & 0     & 0.01  & $10^{-6}$ & 0         & 0.57      & 0.05 \\
Fig.~\ref{fig:flow_I_to_V}(d)                      & 0.03 & 0     & 0.01  & $10^{-6}$ & 0         & 0.67      & 0.05 \\
Fig.~\ref{fig:flow_I_to_V}(e)                      & 0.03 & 0     & 0.01  & $10^{-6}$ & 0         & 0.76      & 0.05 \\
\hline

Fig.~\ref{fig:PD_3D}(a)                            & --   & 0     & 0.01  & --        & 0         & --       & 0    \\
Fig.~\ref{fig:PD_3D}(b)                            & --   & 0.01  & 0.01  & --        & 0         & --       & 0    \\
Fig.~\ref{fig:PD_3D}(c)                            & 0.03 & --    & 0.01  & --        & 0         & --       & 0    \\
\hline

Fig.~\ref{fig:PD_3D_Nmzm}(a)                       & --   & 0     & 0.01  & --        & 0         & --       & 0    \\
Fig.~\ref{fig:PD_3D_Nmzm}(b)                       & --   & 0.01  & 0.01  & --        & 0         & --       & 0    \\
Fig.~\ref{fig:PD_3D_Nmzm}(c)                       & 0.03 & --    & 0.01  & --        & 0         & --       & 0    \\
\hline

Fig.~\ref{fig:K_D_densityplot}(a)                  & 0.03 & 0.01  & 0.01  & $10^{-6}$ & 0         & --        & 0    \\
Fig.~\ref{fig:K_D_densityplot}(b)                  & 0.03 & 0.01  & 0.01  & --        & 0         & 0.6       & 0    \\
\hline
Fig.~\ref{fig:pointPprimeflow}(a)                  & 0.03 & 0.015 & 0.01  & $10^{-5}$ & 0         & 0.7       & 0    \\ 
Fig.~\ref{fig:pointPprimeflow}(b--f)               & 0.03 & --    & 0.01  & --        & 0         & 0.7       & 0    \\ 
\hline
Fig.~\ref{fig:PD_transport2}(a)                    & 0.03 & --    & 0.01  & --        & 0         & 0.6       & 0    \\ 
Fig.~\ref{fig:PD_transport2}(b)                    & 0.03 & --    & 0.005 & --        & 0         & 0.6       & 0    \\ 
Fig.~\ref{fig:PD_transport2}(c)                    & 0.03 & --    & 0.01  & --        & 0         & 0.7       & 0    \\ 
\hline
Fig.~\ref{fig:PD2_Nmzm}(a)                         & 0.03 & --    & 0.01  & --        & 0         & 0.6       & 0    \\ 
Fig.~\ref{fig:PD2_Nmzm}(b)                         & 0.03 & --    & 0.005 & --        & 0         & 0.6       & 0    \\ 
Fig.~\ref{fig:PD2_Nmzm}(c)                         & 0.03 & --    & 0.01  & --        & 0         & 0.7       & 0    \\ 
\hline
Fig.~\ref{fig:asymm_D_transport_PD}                & 0.03 & 0     & 0.01  & --        & $10^{-6}$ & --        & 0    \\
\hline
Fig.~\ref{fig:asymm_D_Nmzm_PD}                     & 0.03 & 0     & 0.01  & --        & $10^{-6}$ & --        & 0    \\
\hline

Fig.~\ref{fig:Kflow}(a)                            & 0.03 & 0     & 0.01  & $10^{-8}$ & 0         & 0.2, 0.4, & 0    \\
                                                   &      &       &       &           &           & 0.6, 0.8  &      \\
Fig.~\ref{fig:Kflow}(b)                            & 0.03 & 0     & 0.01  & $10^{-4}$ & 0         & 0.5, 0.7, & 0    \\
                                                   &      &       &       &           &           & 0.85      &      \\
\hline

Fig.~\ref{fig:flow_Ppp}                            & 0.03 & 0     & 0.01  & $10^{-3}$ & 0         & 0.8       & 0    \\ 
\hline
Fig.~\ref{fig:backscattering_induced_Nmzm1}(a)     & 0.03 & --    & 0.01  & --        & 0         & 0.75      & 0    \\
Fig.~\ref{fig:backscattering_induced_Nmzm1}(b)     & 0.03 & --    & 0.01  & --        & 0         & 0.8       & 0    \\
\hline 
Fig.~\ref{fig:m3_PD_transport}~\footnotemark[1]    & 0.03 & 0     & 0.01  & --        & 0         & --        & 0    \\ 
\hline
Fig.~\ref{fig:m3_flow_Ppp}~\footnotemark[2]        & 0.03 & 0     & 0.01  & $10^{-3}$ & 0         & 0.8       & 0    \\ 
\hline
Fig.~\ref{fig:m3_flow_Pppp}                        & 0.03 & 0     & 0.01  & $10^{-3}$ & 0         & 0.3       & 0    \\

\hline
Fig.~\ref{fig:mzm_density_channel}(a)           & 0.03 & 0.01  & 0.01  & $10^{-6}$ & 0         & 0.54      & 0    \\
Fig.~\ref{fig:mzm_density_channel}(b)              & 0.03 & 0.01  & 0.01  & $10^{-6}$ & 0         & 0.57      & 0    \\
Fig.~\ref{fig:mzm_density_channel}(c)              & 0.03 & 0.01  & 0.01  & $10^{-6}$ & 0         & 0.61      & 0    \\
\hline \hline
\end{tabular} 
\footnotetext[1]{The parameter sets used here are identical to those in Fig.~\ref{fig:PD_transport1}(a), except for $m=3$ in the former.}
\footnotetext[2]{The parameter sets used here are identical to those in Fig.~\ref{fig:flow_Ppp}, except for $m=3$ in the former.}
\label{Table:Parameters}
\end{table}

\section{Details about the numerical analysis}
\label{Appendix:more}

In this section, we list the adopted model parameters in this work and provide more numerical results.

\subsection{Adopted model parameters }

In Table~\ref{Table:Parameters} we list the adopted values of the parameters in the plots throughout the article.

\begin{figure}[th]
\centering
    \includegraphics[width=1.0\linewidth]{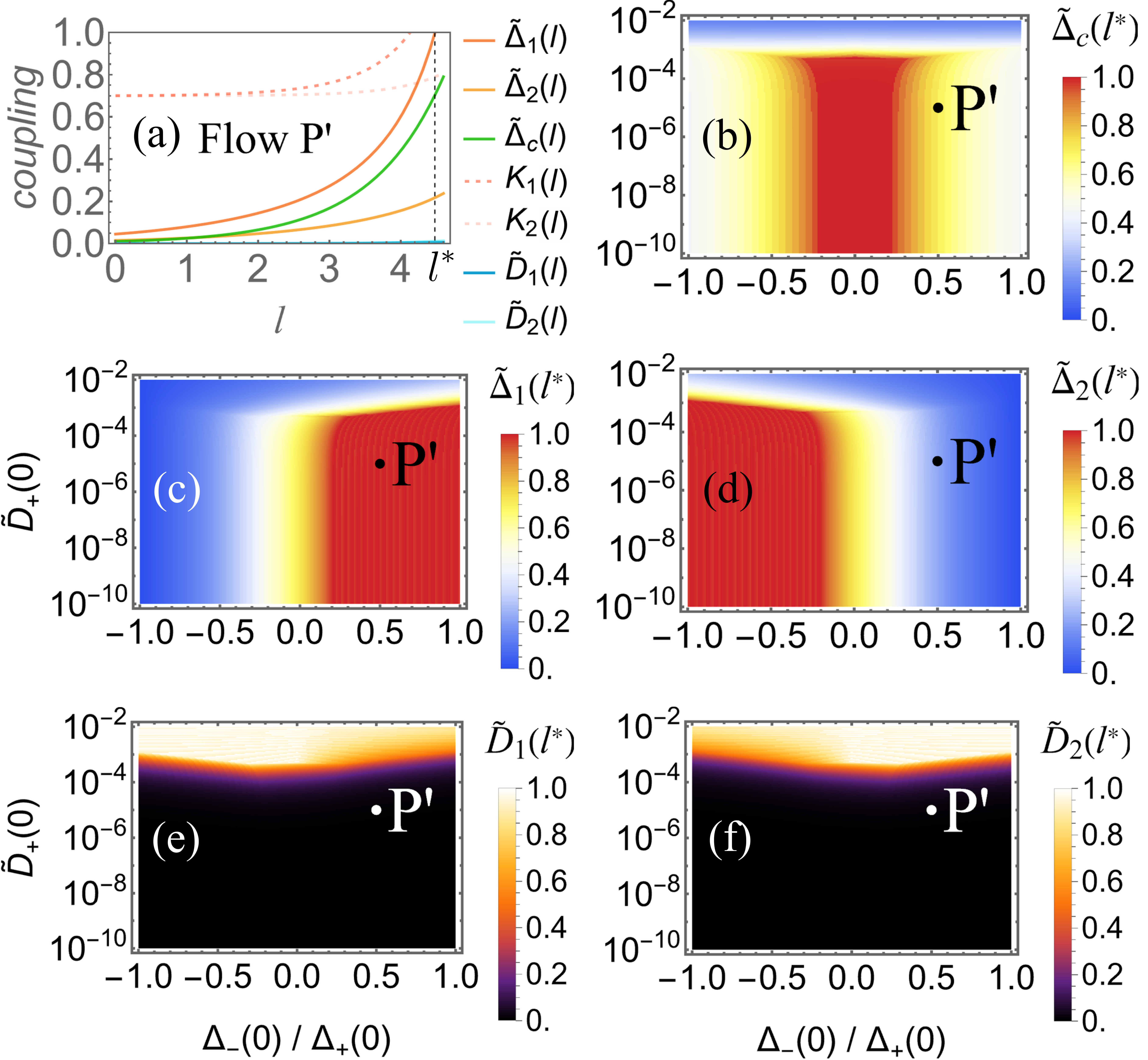}
    \caption{Similar plots to Fig.~\ref{fig:pointPflow} but with 
     color maps in the $[ \Delta_-(0)/ \Delta_+(0),\,\tilde D_+(0) ]$ plane with $\tilde \Delta_+(0) =0.03$ and $K (0) = 0.7$.
    For the  parameter set P$'$, we additionally set $\tilde D_+(0) = 10^{-5}$ and $\tilde \Delta_{-} (0) = 0.015 $. 
    See Table~\ref{Table:Parameters} for the complete set of the adopted parameter values. 
   }
    \label{fig:pointPprimeflow}
\end{figure}

\subsection{Additional  flow example }

In this section, we provide a flow example in Fig.~\ref{fig:pointPprimeflow}, accompanied by a scan of the renormalized coupling strengths through the initial pairing asymmetry and disorder strength. At the representative set P$'$ in Fig.~\ref{fig:pointPprimeflow}(a), the local pairing $\tilde \Delta_1$ is the most relevant, reaching strong coupling first, while the nonlocal pairing $\tilde \Delta_c$ remains subdominant and $\tilde \Delta_2$ is only weakly enhanced. This structure is reflected in the saturation pattern of the overall diagrams in Figs.~\ref{fig:pointPprimeflow}(b--d): $\tilde \Delta_1(l^*)$ dominates at P$'$, $\tilde \Delta_c(l^*)$ is subdominant, and $\tilde \Delta_2(l^*)$ remains weak. The renormalized backscattering couplings in Figs.~\ref{fig:pointPprimeflow}(e--f) stay irrelevant, consistent with the observation in Fig.~\ref{fig:pointPprimeflow}(a). Thus, in contrast to the set P in Fig.~\ref{fig:pointPflow}, where nonlocal pairing takes over, the set P$'$ illustrates a regime where a local pairing prevails.

\begin{figure*}[ht]
\centering
\includegraphics[width=1.0\linewidth]{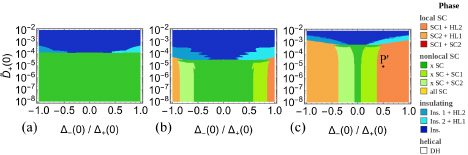}
\caption{Phase diagrams similar to Fig.~\ref{fig:PD_transport1} but in the $ [ \Delta_-(0)/  \Delta_+(0),\,\tilde D_+(0) ]$ plane.
The adopted parameter values include (a) $\tilde \Delta_c(0)=0.01$ and $K_+(0)=0.6$, (b)  $\tilde \Delta_c(0) = 0.005$ and $K_+(0)=0.6$, and (c) $\tilde \Delta_c(0)=0.01$ and $K_+(0)=0.7$.  
See Table~\ref{Table:Parameters} for the complete set of the adopted parameter values.  
}
\label{fig:PD_transport2}
\end{figure*}

\begin{figure*}[ht]
\centering
\includegraphics[width=1.0\linewidth]{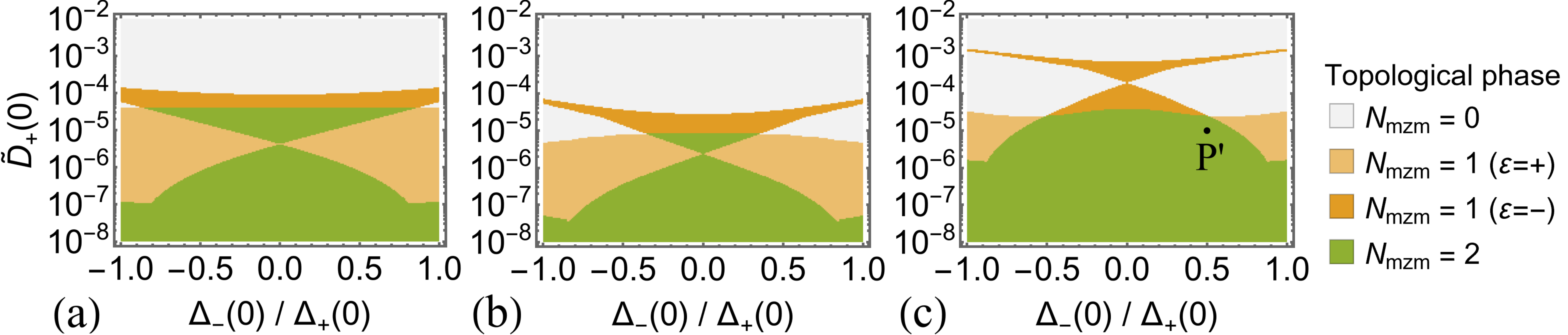}
    \caption{Topological phase diagrams similar to Fig.~\ref{fig:PD1_Nmzm} but in the $[ \Delta_-(0)/ \Delta_+(0),\,\tilde D_+(0) ]$ plane.
    The parameter values include (a)  $\tilde \Delta_c(0)=0.01$ and $K_+(0)=0.6$, (b)  $\tilde \Delta_c(0) = 0.005$ and $K_+(0)=0.6$, and (c) $\tilde \Delta_c(0) = 0.01$ and $K_+(0)=0.7$.       
    See Table~\ref{Table:Parameters} for the complete set of the adopted parameter values.  
    }
    \label{fig:PD2_Nmzm}
\end{figure*}

\subsection{Additional phase diagrams with varying backscattering strength and pairing asymmetry}

In this section, we present phase diagrams obtained by varying the backscattering strength and pairing asymmetry. While the main text focuses primarily on the parameter space spanned by the disorder strength and interaction parameters, here we examine how the transport and topological properties evolve under these additional variations. The transport phase diagrams are discussed in Sec.~\ref{subsubsec:PD_Deltaminus_Dplus}, followed by the corresponding topological phase diagrams in Sec.~\ref{subsubsec:topo_Deltaminus_Dplus}.

\subsubsection{Transport-based phase diagrams}
\label{subsubsec:PD_Deltaminus_Dplus}

In this section, we explore the effects of pairing asymmetry in more detail.
To this end, we investigate the phase diagrams in the $[ \Delta_-(0)/ \Delta_+(0),\,\tilde D_+(0) ]$ plane. An example is presented in Fig.~\ref{fig:PD_transport2}(a). For the chosen interaction strength $K_+(0)=0.6$, the crossed Andreev pairing $ \Delta_c$ dominates over almost the entire diagram: even substantial asymmetry in the local pairings leaves the system in the nonlocal superconducting phase as long as backscattering remains weak. For stronger backscattering strengths, this phase is destroyed and the system becomes insulating. With increasing pairing asymmetry, the insulating region then forms two intermediate phases where backscattering localizes one channel while the other stays metallic, an interesting pattern that will reappear in the following panels.

Furthermore, we show that the effects of the pairing asymmetry depend also on other parameters. In Fig.~\ref{fig:PD_transport2}(b), we consider a reduced initial crossed Andreev pairing, $\tilde \Delta_c(0)=0.005$, with the other parameters identical to Fig.~\ref{fig:PD_transport2}(a). The nonlocal superconducting phase still emerges near the symmetric point, yet it is now more fragile: for a large $|\Delta_{-}(0)|$, the system is driven into local superconducting phases. Between these and small $\Delta_-$ region, intermediate regimes appear in which nonlocal and local superconductivity coexist. The critical backscattering needed to destroy nonlocal pairing is slightly lower than in  Fig.~\ref{fig:PD_transport2}(a), and the channel-selective insulating phases broaden as $|  \Delta_- (0) /  \Delta_+ (0) |$ increases. Thus, with a weaker $  \Delta_c(0)$, strong backscattering is able to pin down one channel over a wider span of initial conditions, leaving the other gapless.

In addition to nonlocal pairing, we find that a weaker interaction can also enhance the pairing asymmetry effects. 
Fig.~\ref{fig:PD_transport2}(c) corresponds to the parameter set with $K_+(0)=0.7$ and otherwise identical to Fig.~\ref{fig:PD_transport2}(a). In this regime, the nonlocal superconductivity survives only in a narrow region around perfect pairing symmetry. Even a small asymmetry is sufficient for the local pairing on one channel to dominate, producing wide local superconducting phases, with intermediate domains with coexisting local and nonlocal superconductivity. Meanwhile, the insulating behavior appears only when the bare backscattering strength is very large. Consistent with Fig.~\ref{fig:PD_transport1}, weakening interactions enhances stability of the superconducting phases while making the system more resilient to backscattering.


\subsubsection{Topological phase diagrams}

\label{subsubsec:topo_Deltaminus_Dplus}

We now present a detailed analysis of the effects of pairing asymmetry on the topological phase diagrams, as shown in Fig.~\ref{fig:PD2_Nmzm}. The overall phase diagram is symmetric with respect to the $\tilde {\Delta}_{-}(0) = 0$ axis. 
At a fixed $K_+(0)$ in Fig.~\ref{fig:PD2_Nmzm}(a), as $|\tilde \Delta_-(0)|$ increases, the emerging topological regions with $N_{\rm mzm}=1$ become wider.  

In Fig.~\ref{fig:PD2_Nmzm}(b), we consider a weaker initial nonlocal pairing. 
The nontrivial region becomes noticeably smaller, which follows directly from Eq.~\eqref{eq:inequality_symm_dis}: as $\tilde \Delta_c(l^{*})$ weakens, the upper bound of the inequality is lowered, shrinking the nontrivial region. On the other hand, the overall shape of the $\varepsilon=-$ boundary changes very little. This is because the corresponding bound, $[\Delta_+(l^{*})-  V_+(l^{*})]$, is only weakly affected by $\Delta_c $, in contrast to the $\varepsilon=+$ branch.

Figure~\ref{fig:PD2_Nmzm}(c) shows the same parameter scan as Fig.~\ref{fig:PD2_Nmzm}(a) but at a weaker fixed interaction strength. In this case, the $N_{\mathrm{mzm}} = 2$ region expands over a broader range at low backscattering strengths. As expected, since backscattering is less relevant in this interaction regime, its influence on the topological condition and on the associated phase transitions [cf. Fig.~\ref{fig:PD1_Nmzm}(c)] is reduced, leaving the $N_{\mathrm{mzm}} = 2$ phase more extended in the diagram.

\subsection{RG flow examples for $\tilde D_-(0)\ne 0$ }
\label{Appendix:RG_disorder_asymmetry}

In this section we examine how a slight asymmetry in the bare backscattering strengths, implemented by setting $\tilde D_-(0)=10^{-6}$, modifies the transport and topological phase diagrams while keeping all the other parameters identical to those in Fig.~\ref{fig:PD_transport1}(a).

\begin{figure}[t]
    \centering
    \includegraphics[width=0.9\linewidth]{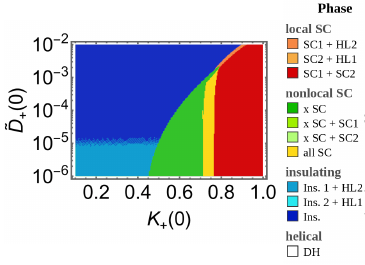}
    \caption{Phase diagram in the $[ K_+(0),\,\tilde D_+(0) ]$ plane for the parameter set used in Fig.~\ref{fig:PD_transport1}(a) but with $\tilde D_-(0) = 10^{-6}$. See Table~\ref{Table:Parameters} for the complete set of the adopted parameter values.}\label{fig:asymm_D_transport_PD}
\end{figure}

As shown in Fig.~\ref{fig:asymm_D_transport_PD}, the transport phase diagram does not show visible difference compared to the symmetric case in the main text. 
In particular, all superconducting phases appear in essentially the same regions, indicating that a disorder imbalance of the order $10^{-6}$ has little to no effect on the competition among the pairing processes.
On the other hand, a notable difference emerges only in the regime of small $\tilde D_+(0)$. The fully insulating phase changes into a regime where only one channel localizes while the other remains a helical liquid.
This occurs because when the order of magnitude of $\tilde D_+(0)$ becomes comparable to the fixed asymmetry $\tilde D_-(0)=10^{-6}$, the bare backscattering strengths $\tilde D_{1,2}(0) $ differ appreciably. The channel with the larger initial backscattering strength then flows to strong coupling more rapidly, dominating the flow. 

\begin{figure}[t]
    \centering \includegraphics[width=0.95\linewidth]{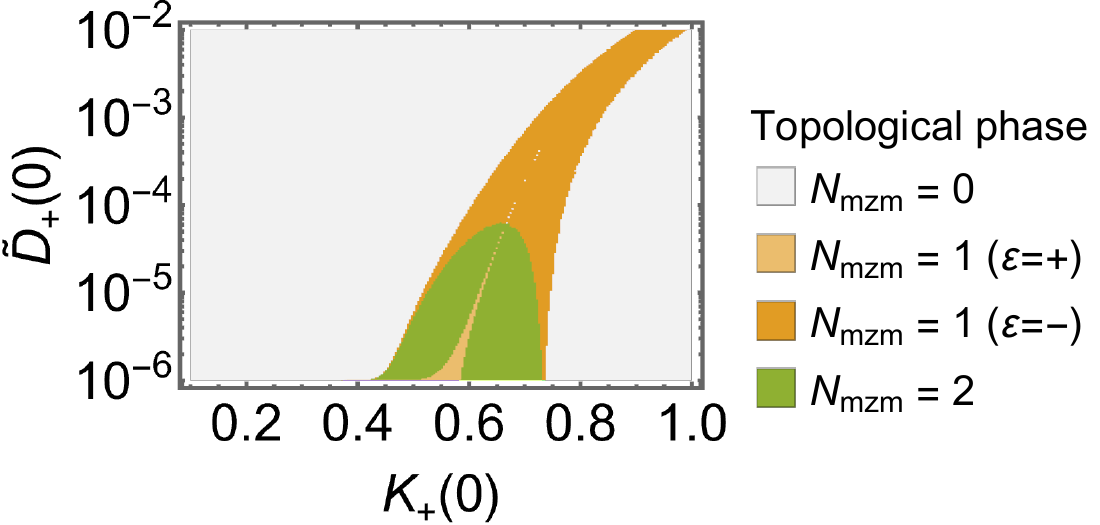}
    \caption{Topological phase diagram for the parameter set used in Fig.~\ref{fig:PD1_Nmzm}(a) but with $\tilde D_-(0) = 10^{-6}$. See Table~\ref{Table:Parameters} for the complete set of the adopted parameter values. }
    \label{fig:asymm_D_Nmzm_PD}
\end{figure}

We also examine the topological phase diagram and show the results in Fig.~\ref{fig:asymm_D_Nmzm_PD}.
In addition to the overall reduction of the topological region, there is also a development of an additional $N_{\rm mzm}=1$ phase that separates the two regions with $N_{\rm mzm}=2$. 
As in the main text, the Majorana zero mode corresponding to the $\varepsilon=-$ branch vanishes inside this emerging region. 
Interestingly, this structure closely mirrors a similar behavior in the system in the presence of pairing or Coulomb asymmetry in  Figs.~\ref{fig:PD1_Nmzm}(b--c) in the main text.
We conclude this section by noting that all the three types of asymmetry: pairing, interaction, and now disorder, lead to emergence of additional topological phases in terms of $N_{\rm mzm}$.

\subsection{Effective model with refermionization }
\label{Appendix:refermionization}

\begin{figure}[t]
    \centering
    \includegraphics[width=0.95\linewidth]{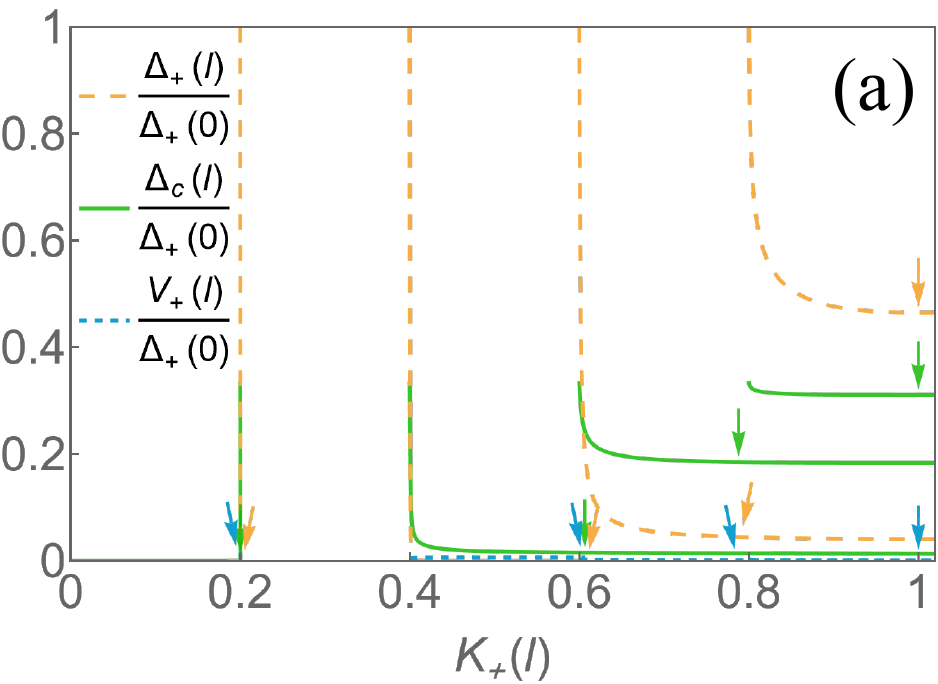}
    \includegraphics[width=0.95\linewidth]{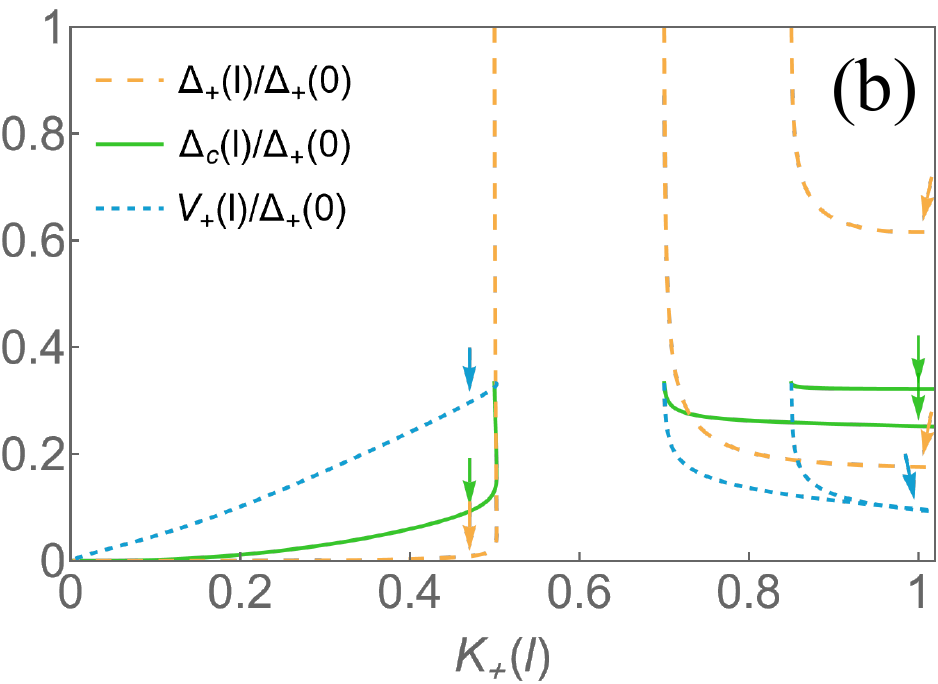}
    \caption{Representative RG flow trajectories for (a) $\tilde D_+(0)=10^{-8}$ and 
    (b) $\tilde D_+(0)=10^{-4}$.
    The arrows indicate the RG flow stopping points and the other parameters include  $\tilde \Delta_+(0)=0.03,\,\tilde \Delta_c=0.01$, and $\tilde \Delta_-(0)=\tilde D_-(0)=K_-(0)=0$; see Table~\ref{Table:Parameters} for the complete set of the adopted parameter values.  
    }
    \label{fig:Kflow}
\end{figure}

In this section we demonstrate how an effective model with the single-particle description can be derived from the refermionization procedure.
To this end, we show two representative RG flow examples in Fig.~\ref{fig:Kflow}. For simplicity, we focus on symmetric cases, $\tilde \Delta_-(0)=\tilde D_-(0)=K_-(0)=0$, here.
We note that the presence of   asymmetries studied in the main text does not modify the qualitative behavior.

As shown in Fig.~\ref{fig:Kflow}(a), when the initial backscattering $\tilde D_+(0)$ is sufficiently small, the RG flow trajectories  can be adiabatically connected to the noninteracting limit without closing the system gap. 
For broader parameter regime, Fig.~\ref{fig:Kflow}(b) shows that the same procedure still applies for large initial backscattering, provided that the interaction is weak. Although the backscattering term is now comparable to the nonlocal pairing at short scales, the trajectories still enter the superconducting regime before disorder becomes dominant. For both panels in Fig.~\ref{fig:Kflow}, the couplings beyond the stopping scale $l^*$ remain constant upon evolving to the noninteracting point $K_+=1$.

The above observation allows us to adiabatically connect the RG flow to the noninteracting limit, at which we can refermionize the system to obtain the effective model with renormalized coupling strengths at $l^*$.
 Utilizing the effective model, we then identify the topological character of various superconducting phases, as discussed in the main text. 

For completeness, we also remark on the opposite behavior in the regime of strong interaction and disorder.
In this regime, the interaction parameters are driven downward, as indicated in the RG flow equation~\eqref{eqn:rg_flow_K} and they do not pass through the noninteracting limit. In this regime, one obtains a trivial insulating phase, and the refermionization procedure is unnecessary.

\subsection{Details about the disorder-induced topological phase transitions }
\label{Appendix:more_on_Nmzm1}

In this section, we provide more details about the disorder-induced topological phase transitions discussed in Sec.~\ref{subsubsec:topo_Nmzm1}.

\begin{figure}[ht]
\centering
\includegraphics[width=0.85\linewidth]{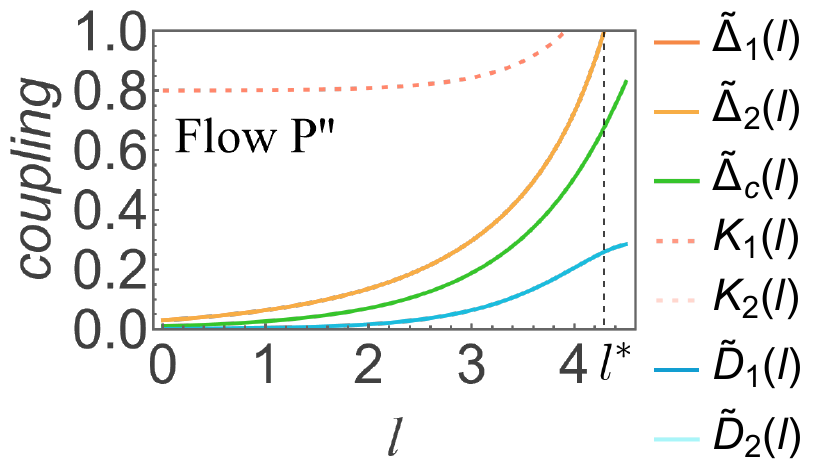}
\caption{RG flow  for the parameter set P$''$ in Fig.~\ref{fig:PD1_Nmzm}(a).
Note that $\tilde{\Delta}_1$ and $\tilde{\Delta}_2$ flow identically, and thus their curves overlap in the plot. The same applies to $\tilde{D}_1$ and $\tilde{D}_2$.
See Table~\ref{Table:Parameters} for the complete set of the adopted parameter values. 
}
    \label{fig:flow_Ppp}
\end{figure}

In Fig.~\ref{fig:flow_Ppp}, we show the RG flow with the initial set labeled as P$''$ in Fig.~\ref{fig:PD1_Nmzm}.
The result shows identical RG flow and renormalized couplings in the two channels. Therefore, we note that the $N_{\rm mzm} = 1$ region is not due to flow to distinct channels, but a consequence of topological criterion shifted by the presence of disorder.

\begin{figure}[t]
\centering
\includegraphics[width=1.0\linewidth]{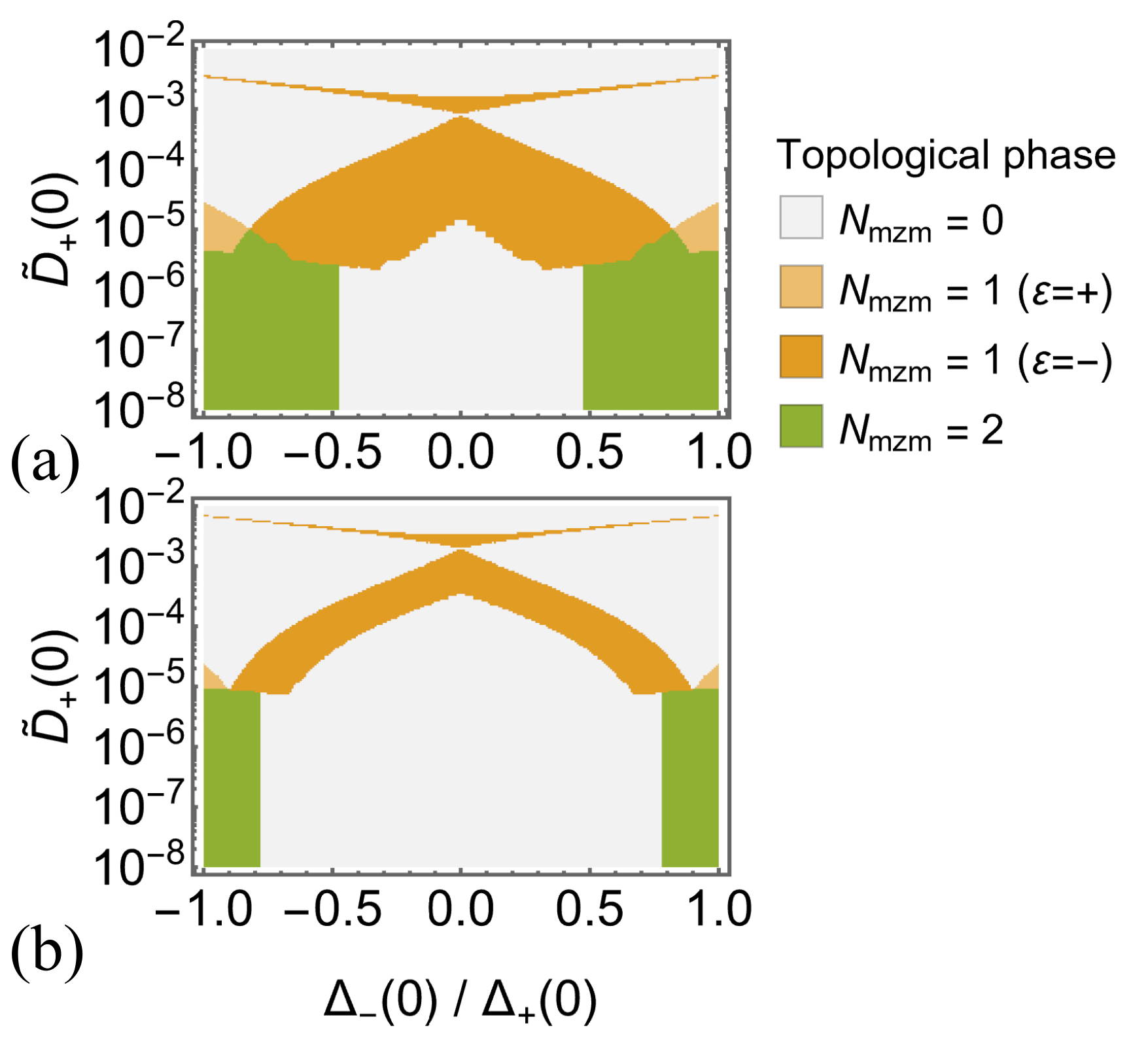}
    \caption{Topological phase diagrams similar to Fig.~\ref{fig:PD2_Nmzm}.
    The adopted values of the parameters are the same as those in Fig.~\ref{fig:PD2_Nmzm}(a,c), but with (a) $K_+(0)=0.75$ and (b) $K_+(0)=0.8$.
    See Table~\ref{Table:Parameters} for the complete sets of the adopted parameter values.   
    }
    \label{fig:backscattering_induced_Nmzm1}
\end{figure}

In Fig.~\ref{fig:backscattering_induced_Nmzm1}, we show additional phase diagrams in alignment with disorder-induced topological phase transitions discussed in Sec.~\ref{subsubsec:topo_Nmzm1}.
As shown in Fig.~\ref{fig:backscattering_induced_Nmzm1}(a), we examine the dependence on pairing asymmetry at $K_+(0) = 0.75$. While the clean system is topologically trivial at $\tilde{D}_+(0) = 0$, increasing the backscattering strength drives the system into the $N_{\mathrm{mzm}} = 1$ phase. Figure~\ref{fig:backscattering_induced_Nmzm1}(b) shows the same scan at a slightly weaker interaction, $K_+(0) = 0.8$, where the $N_{\mathrm{mzm}} = 1$ region becomes noticeably narrower.

\begin{figure}[h]
    \centering
    \includegraphics[width=0.95\linewidth]{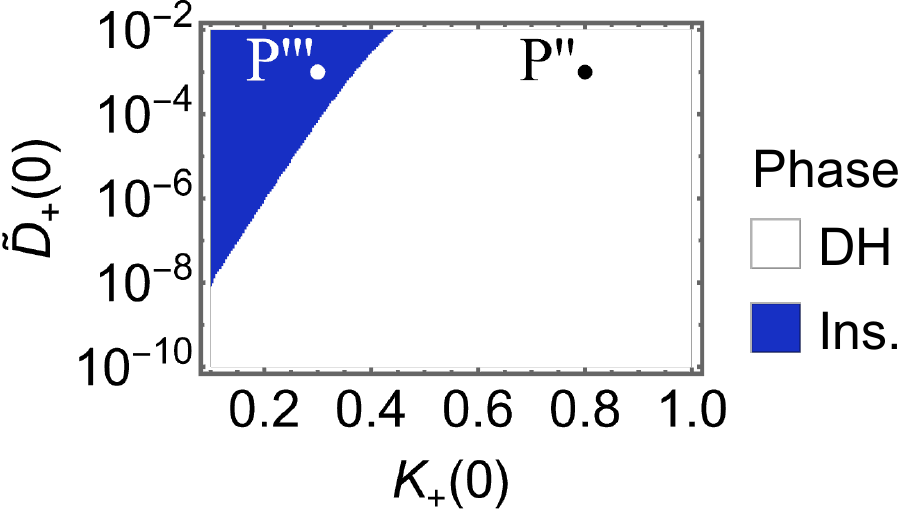}
    \caption{Phase diagram for the parameter set used in Fig.~\ref{fig:PD_transport1}(a) but with $m=3$. The dots P$''$ and P$'''$ mark representative sets of initial conditions used in the RG flow plots in Figs.~\ref{fig:m3_flow_Ppp}--\ref{fig:m3_flow_Pppp} below. See Table~\ref{Table:Parameters} for the complete set of the adopted parameter values.}
    \label{fig:m3_PD_transport}
\end{figure}

\subsection{Numerical results for fractional helical liquids }
\label{Appendix:m3_fractional}

In this section, we briefly discuss how the above results are modified for fractional helical liquids with $m>1$ for completeness.
The RG flow equations in Sec.~\ref{sec:rg_flow} show that increasing $m$ reduces the relevance of the couplings, since the scaling exponents of $\tilde \Delta_n$, $\tilde \Delta_c$, and $\tilde D_n$ all increase with $m$ at fixed $K_n$. 

\begin{figure}[t]
\centering\includegraphics[width=0.9\linewidth]{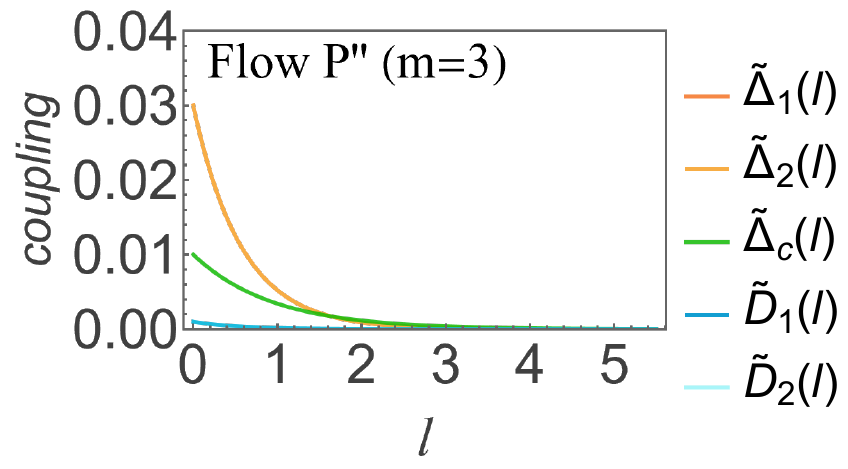}
    \caption{RG flow of the couplings for the parameter set P$''$ in Fig.~\ref{fig:m3_PD_transport}; the corresponding $m=1$ RG flow is shown in Fig.~\ref{fig:flow_Ppp}. Note that $\tilde{\Delta}_1$ and $\tilde{\Delta}_2$ flow identically, and thus their curves overlap in the plot. The same applies to $\tilde{D}_1$ and $\tilde{D}_2$. See Table~\ref{Table:Parameters} for the complete set of the adopted parameter values. 
    }
    \label{fig:m3_flow_Ppp}
\end{figure}

\begin{figure}[ht]
    \centering
    \includegraphics[width=0.9\linewidth]{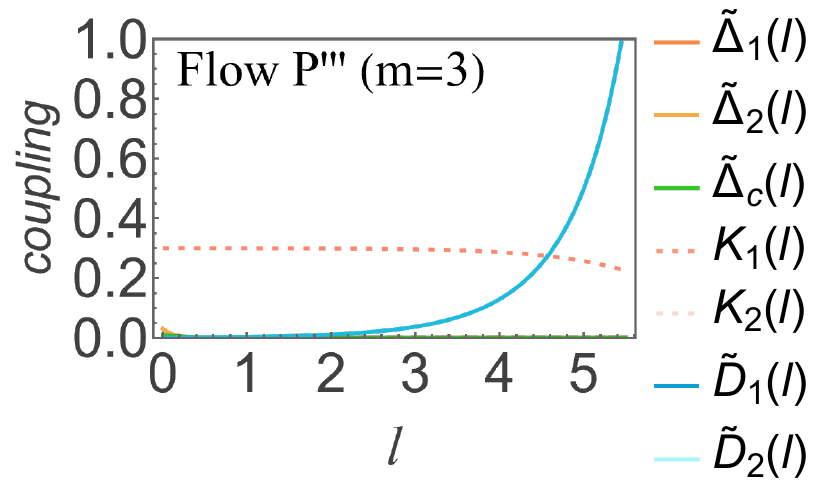}
    \caption{Similar plot to Fig.~\ref{fig:m3_flow_Ppp} but for the parameter set P$'''$ marked in Fig.~\ref{fig:m3_PD_transport}. See Table~\ref{Table:Parameters} for the complete set of the adopted parameter values. 
    }
    \label{fig:m3_flow_Pppp}
\end{figure}

As an illustrative example, we consider $m=3$, corresponding to fractional quantum spin Hall edges at $1/3$-filling, and keep the same parameters as in Fig.~\ref{fig:PD_transport1}(a) in the main text. 
The results are presented in Fig.~\ref{fig:m3_PD_transport},
which extends the RG results form the integer quantum spin Hall edge in Fig.~\ref{fig:flow_Ppp}. In this fractional case, all pairing couplings become irrelevant and therefore the superconductivity vanishes in the diagram. The relevance of backscattering is also reduced, as reflected by the noticeably smaller insulating region. Nevertheless, strong initial repulsion together with sufficiently large bare backscattering still drives the system into the insulating phase. 
We select two representative sets, P$''$ in the double-helical-liquid region and P$'''$ in the insulating region, and examine their RG flow trajectories in Figs.~\ref{fig:m3_flow_Ppp}--\ref{fig:m3_flow_Pppp}.

For the representative point P$''$ in the metallic region, backscattering remains irrelevant throughout the flow, and all the couplings decrease with $l$, as shown in Fig.~\ref{fig:m3_flow_Ppp}.  
For the other set, P$'''$, the stronger e-e interaction, $K_+(0)=0.3$, places the system inside the insulating region of Fig.~\ref{fig:m3_PD_transport}. 
As shown in Fig.~\ref{fig:m3_flow_Pppp}, the superconducting couplings remain irrelevant throughout the flow, while the backscattering terms become sufficiently relevant to grow rapidly and reach order unity. This drives the system into the nonsuperconducting phases owing to the increased scaling dimensions at $m=3$.
In any case, we see that it is difficult to maintain the proximity-induced pairing in the fractional edges; we thus focus on the integer quantum spin Hall edge in the main text.

\subsection{Representative spatial density profiles of the zero modes}

\begin{figure}[h]
    \centering
    \includegraphics[width=1.0\linewidth]{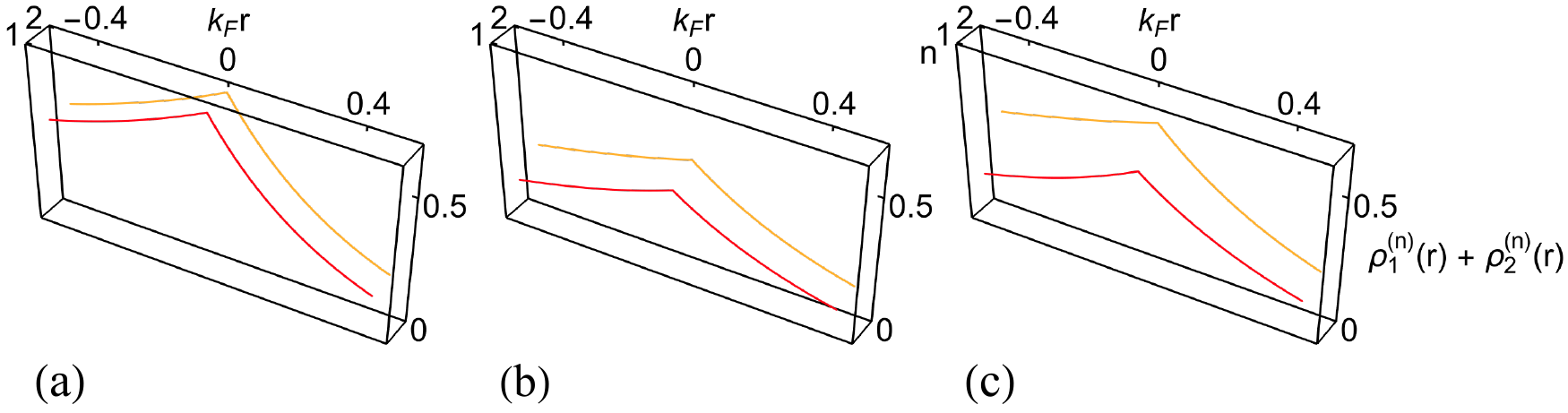}
    \caption{Channel-resolved density profiles $\rho^{(n)}(r)\equiv \sum_{j} |\Psi_{\text{mzm},j}^{(n)}(r)|^2$ of the Majorana zero modes along the local coordinate $r$.
    From (a) to (c), we have $K_+(0)=0.54,\,0.57$ [identical to the point P  in Fig.~\ref{fig:PD1_Nmzm}(b)] and $0.61$, respectively. The red (orange) curves show the density for the channel $n=1$ ($n = 2$). 
    The other parameters include $\tilde \Delta_+(0)=0.03$, $\tilde\Delta_-(0)=0.01$, $\tilde\Delta_c(0)=0.01$, $\tilde D_+(0)=10^{-6}$ and $\tilde D_-(0)=K_-(0)=0$; see Table~\ref{Table:Parameters} for the complete sets of the adopted parameter values. 
    }
    \label{fig:mzm_density_channel}
\end{figure}

To demonstrate the spatial density profile of the decaying modes, we show the numerical results for three representative values of $K_+(0)$ in Fig.~\ref{fig:mzm_density_channel} and plot the channel resolved density profiles.
Figs.~\ref{fig:mzm_density_channel}(a--c) show the resulting channel-resolved density profiles for the corresponding values of $K_+(0)$. Figs.~\ref{fig:mzm_density_channel}(a,c) correspond to the $N_{\rm mzm}=2$ regions, while Fig.~\ref{fig:mzm_density_channel}(b) is in the intermediate region where one of the modes vanishes, $N_{\rm mzm}=1$. In the resulting plots, we observe exponentially decaying amplitudes. Consistent with the analysis in the main text, we note sharper localization on the $r>0$ side and slower decay on the other side. We can also observe broadening of the profiles as the interactions get weaker.

\end{document}